\documentclass{article}
\usepackage[english]{babel}
\usepackage[T1]{fontenc}
\usepackage{fullpage}
\usepackage[utf8]{inputenc}
\usepackage{newlfont}
\usepackage{fancyhdr}
\usepackage{mhchem}
\usepackage{indentfirst}
\usepackage{graphicx}
\usepackage{newlfont}
\usepackage{soul}
\usepackage{amssymb}
\usepackage{amsmath}
\usepackage{mathtools}
\usepackage{latexsym}
\usepackage{amsthm}
\usepackage{cite}
\usepackage{bm}
\usepackage{bbold}
\usepackage{xcolor}
\usepackage{cancel}
\usepackage{listings}
\usepackage[hidelinks]{hyperref}
\usepackage[]{cleveref}
\usepackage{subfigure}
\newcommand{\rH}{r_{\text{H}}}
\newcommand{\lP}{\ell_{\text{P}}}
\newcommand{\RS}{R_{\text{S}}}
\newcommand{\rst}{r_\ast}
\newcommand{\omegaI}{\omega_\text{I}}
\newcommand{\omegaR}{\omega_\text{R}}

\makeatletter
\newcommand{\mathleft}{\@fleqntrue\@mathmargin0pt}
\newcommand{\mathcenter}{\@fleqnfalse}
\makeatother

\begin{document}
\title{\bf Effective models  of  non-singular quantum black holes}
\author{M.~Cadoni${}^{ab}$\thanks{E-mail: mariano.cadoni@ca.infn.it}, \ M.~Oi${}^{ab}$\thanks{E-mail: mauro.oi@ca.infn.it}, \ 
A. P. ~Sanna${}^{ab}$\thanks{E-mail: asanna@dsf.unica.it} \ 
\\
${}^a$\emph{Dipartimento di Fisica, Universit\`a di Cagliari}
\\
{\em Cittadella Universitaria, 09042 Monserrato, Italy}
\\
\\
${}^b$\emph{I.N.F.N, Sezione di Cagliari}
\\
{\em  Cittadella Universitaria, 09042 Monserrato, Italy}
\\
\\}

\maketitle
\begin{abstract}
We investigate how the resolution of the singularity problem for the  Schwarzschild black hole could be related to the presence of  quantum gravity effects at horizon scales. Motivated by the analogy with the cosmological Schwarzschild-de Sitter solution, we construct a broad class of non-singular, static, asymptotically-flat black-hole solutions with a de Sitter (dS) core, sourced by an anisotropic fluid, which effectively encodes the quantum corrections. The latter are parametrized by a single length-scale  $\ell$,  which has a dual interpretation as an  effective "quantum hair"  and   as the length-scale resolving the classical singularity. Depending on the value of $\ell$, these  solutions  can  have  two   horizons, be extremal (when the two horizons merge) or  be horizonless exotic stars. We also investigate the thermodynamic behavior of our black-hole solutions and propose a generalization of the area law in order to account for their entropy.   We find  a second-order phase transition near extremality, when $\ell$ is of order of the classical Schwarzschild radius $\RS$. Black holes with $\ell\sim \RS$   are thermodynamically  preferred  with respect to those with $\ell\ll\RS$,  supporting  the relevance of quantum corrections  at   horizon scales. We also find that the extremal configuration is a zero-temperature, zero-entropy state with its  near-horizon geometry factorizing as AdS$_2 \times$ S$^2$, signalizing the possible relevance of these models for the information paradox. Finally, we show that the presence of  quantum corrections with  $\ell \sim \RS$  have   observable phenomenological signatures  in  the photon orbits and in the  quasi-normal modes (QNMs) spectrum.
In particular, in the near-extremal regime, the imaginary part of the QNMs spectrum scales with the temperature  as $c_1/\ell+ c_2\ell T_\text{H}^2$, while it goes to zero linearly in the near-horizon  limit. Our general findings are confirmed by revisiting two already-known models, which are  particular cases  of our general class of models,  namely the Hayward and gaussian-core black holes. 
\end{abstract}

\tableofcontents

\section{Introduction}
Since the discovery of the Schwarzschild solution, the presence of a singularity inside black holes, together with the initial cosmological one, has represented a serious challenge to our current understanding of the fundamental laws of physics. This problem became even more serious after the groundbreaking Penrose and Hawking singularity theorems \cite{Penrose:1964wq, Hawking:1970zqf}. They proved incontrovertibly that, under a set of a few, very general and physically motivated assumptions (the validity of the weak energy condition and either global hyperbolicity or the validity of the strong energy condition), these space-time singularities are unavoidable, at least in the classical general relativity (GR) framework. Despite this, it is conjectured that these singularities are always hidden behind a causal barrier, the event horizon, which prevents outside observers from seeing them and the theory from completely loosing its predictive power \cite{Penrose:1969pc}. Semiclassical effects, like black hole evaporation \cite{Hawking:1974sw}, seem however to bring the singularity problem back on the table, as the final steps of the evaporation process, where the singularity role should be most prominent, are still poorly understood. 

Although it is in principle solvable already in the classical GR framework by relaxing some assumptions of Penrose's theorem and constructing non-singular effective models (see, e.g. Refs. \cite{Carballo-Rubio:2018jzw,Carballo-Rubio:2019fnb,Maeda:2021jdc,Simpson:2019mud,Lobo:2020ffi,Mazza:2021rgq,Franzin:2021vnj};  for models  with non-linearly coupled electromagnetic fields, see \cite{Ayon-Beato:1999qin,Bronnikov:2000vy,Dymnikova:2004zc,Culetu:2014lca,Banerjee:2022len,Bokulic:2022cyk}) the singularity problem calls for the need of a quantum description of gravitational interactions. The most widely adopted approach in the past has been to assume that these quantum corrections should only be relevant at Planck scale, $\lP = \sqrt{G}\sim 10^{-35} \ \text{m}$ \footnote{We adopt natural units, $c=\hbar =1$. We will use $G$ or $\lP^2$ interchangeably.} \cite{Tseytlin:1995uq,Lawrence:1995ct,Horowitz:1989bv,Modesto:2004xx, Nicolini:2005vd,Modesto:2008im, Nicolini:2008aj, Hossenfelder:2009fc,Modesto:2010uh, Spallucci:2011rn,Sprenger:2012uc,Bambi:2013caa,Frolov:2014jva,Casadio:2014pia,Binetti:2022xdi}. Quantum gravity effects  should become important only when  the Compton length of a pointlike mass  $M$ becomes comparable with its Schwarzschild radius,  $\RS=2 G M$. Thus, they should be irrelevant as long as gravitational interactions at Planck-scale distances are not considered, like the final stages of the evaporation of black holes, the behavior of space-time near their central singularity or the initial phases of the evolution of the universe.

In recent times, however, there have been many indications supporting the possibility of having relevant quantum gravity effects even at scales much larger than $\lP$, i.e. at horizon or cosmological scales. At the black-hole level, this new perspective gains motivation from different approaches: the firewall paradox \cite{Almheiri:2012rt}, which triggered several recent advances in tackling the black-hole information puzzle (islands and replica wormholes \cite{Penington:2019kki,Almheiri:2019qdq,Almheiri:2020cfm,Bousso:2022ntt}, non-local modifications of effective field theory \cite{Giddings:2012gc,Giddings:2021qas,Giddings:2022jda}, fuzzball proposal \cite{Mathur:2005zp, Mathur:2019dhf,Mathur:2020ely}); 
the emergent gravity and corpuscular gravity scenarios  \cite{Verlinde:2016toy,Dvali:2011aa, Dvali:2013eja, Dvali:2020wqi, Casadio:2015lis, Casadio:2016zpl, Cadoni:2020mgb}, in which a black hole is considered as a coherent state of a large number of gravitons of typical wavelength $\sim \RS$ \cite{Dvali:2010bf,Dvali:2010jz,Dvali:2011th}; finally, the quasi-normal modes (QNMs) spectrum of the Schwarzschild black hole, whose description is consistent with that of an ensemble of oscillators with typical frequency $\omega\sim 1/\RS$ \cite{Maggiore:2007nq,Cadoni:2021jer}. Further evidence came from the galactic and cosmological framework, where deviations from Newtonian dynamics and the evolution of dark energy can be interpreted in terms of long-range quantum gravity effects, described  by an exotic source of Einstein's equation in the form of an anisotropic fluid \cite{Verlinde:2016toy, Cadoni:2018dnd,Cadoni:2017evg, Tuveri:2019zor, Cadoni:2020izk, Cadoni:2020jxe, Cadoni:2021zsl}.

The possibility of having quantum-gravity effects operating at black-hole horizon scales is also extremely interesting from a phenomenological point of view.  These effects  are expected to be encoded in the QNMs spectrum and to be detected by the next generation of gravitational wave (GW) detectors, like the Einstein Telescope (ET), in the ringdown phase of two compact objects merging to form a single black hole. In some particular cases, a manifest signature could be the presence of echoes in the GW signal \cite{Oshita:2019sat, Wang:2019rcf,Chakraborty:2022zlq}.\\

The  starting  idea   of this paper  is   that the resolution of the singularity  problem   could be related to the presence of  quantum gravity effects  at horizon scales.  This  is somehow natural  because we  expect quantum  effects to be  at work both in the smearing  of the classical black-hole singularity and in generating an effective {\sl{quantum hair}} at horizon scales. We parametrize the smearing of the classical singularity with a length-scale $\hat L$, whereas the  quantum hair is represented by an extra length-scale $\ell$. We assume that this smearing is sourced by an exotic form of matter having the form of an anisotropic fluid, which should give an effective description of quantum gravity effects. The analogy  with galactic dynamics, where an infrared (IR) scale $R_0=\sqrt {\RS L}$ is generated out of $\RS$ and the size of the cosmological horizon $L$ \cite{Verlinde:2016toy}, now suggests that, similarly, $\ell$  can be interpreted as an IR scale generated from $\RS$ and $\hat L$, for instance  by the simple relation 
\begin{equation}
  \label{scaling}
\ell=\RS^a\hat L^b
\end{equation}
with $a+b=1$. Thus, the origin of the quantum hair $\ell$ should find explanation in the multi-scale behavior of gravitational interactions.

Following the cosmological analogy, we  can think of a  non-singular black hole as  a "reversed" Schwarzschild-de Sitter (SdS) space-time, in which the external cosmological horizon and the inner Schwarzschild one are interchanged, and for which the length-scale $\hat L$ becomes the de Sitter (dS) length. In this way, we are motivated to construct a general class of non-singular,  static, asymptotically-flat black-hole solutions with a dS core, sourced by an anisotropic fluid, which endows the classical Schwarzschild solution with a quantum hair $\ell$. Extending this similarity with the SdS case and with the dynamically generated scale $R_0$, we will explicitly prove that $\ell$ is dynamically generated by $\RS$ and $\hat L$ by  $\ell \sim \RS^{1/3} \hat L^{2/3}$, a relation which should hold in general for regular models with dS cores.

We find that imposing a regular dS core $a)$ always violates the strong energy condition in the interior of these objects, and therefore allows us to circumvent  the singularity theorem, and $b)$ depending on the value of the parameter $\ell$, our non-singular models can have two, one (extremal configuration) or no horizons. We then proceed by investigating the implications of the presence an extra parameter $\ell$,  assumed to be of the same order of magnitude as $\RS$,  on the thermodynamic properties of the black hole and on the phenomenology of the models, i.e. on photon orbits and on the QNMs spectrum. 

By using the first law of thermodynamics, we show that the presence of $\ell$ causes deviations from the standard area law. We propose therefore an entropy formula  to generalize the  latter. Using this general entropy formula, we also find that the extremal configuration is a zero-temperature, zero-entropy state, a behavior drastically different from extremal Reissner-Nordstr\"om (RN) and Kerr black holes. This, together with the fact that the extremal, near-horizon,  geometry  factorizes  as the  tensor product of two-dimensional Anti de Sitter (AdS$_2$) with a two-sphere, i.e. AdS$_2 \times$ S$^2$, indicates that these regular models could actually be relevant for tackling the information paradox \cite{Hawking:1976ra,Page:1993wv,Mathur:2009hf,Kitaev:2017awl,Almheiri:2019qdq,Almheiri:2019hni}. 
By investigating the behavior of the specific heat and the free energy of the hole, we find a second-order phase transition near extremality, i.e for $ \ell\sim \RS$. In particular, black holes with $\ell\sim \RS$ are energetically preferred  with respect to those with $\RS \gg \ell$, lending further support to the  possible relevance of quantum corrections at horizon scale. 

On the phenomenological side, we find that, for black holes with $\ell \ll \RS$, deviations from standard results concerning   photon orbits and the QNMs spectrum are negligibly small and not detectable, at least in the near future. Conversely, black holes with $\ell \sim \RS$ are characterized  by macroscopic deviations from the Schwarzschild behavior, whose signatures  are potentially detectable by the next generation of GW detectors. In particular, by analytically computing the QNMs spectrum in the eikonal   approximation, we find that, in the near-extremal limit, the imaginary part of the quasi-normal frequencies scale with the black-hole temperature as  $c_1/\ell+ c_2\ell T_\text{H}^2$ (with $c_{1,2}$ constants), while in the near-extremal and near-horizon regimes, it goes to zero, in agreement with several results in the literature \cite{LopezOrtega:2011np, Cordero:2012je, Kettner:2004aw, Bhattacharjee:2020nul, Cadoni:2021qfn,Hod:2008se,Hod:2008zz,Hod:2011zzd,Hod:2012zzb,Hod:2015hga,Zimmerman:2015trm,Joykutty:2021fgj}.  This appears to be a general feature of non-singular black holes, common also to charged and/or rotating extension of regular models \cite{Ansoldi:2006vg,Modesto:2010rv, Lan:2020fmn}.

In the final part of the paper, we check our results by revisiting two already-known models, namely the Hayward and gaussian-core  black-hole metrics, which represent particular cases of our general class of regular black holes. \\

The outline of the paper is the following. In \cref{Sec:2} we build up the grounds for our multi-scale description of gravity by  drawing an analogy between the SdS solution and galactic dynamics from one side, and regular black-hole models for the other side. 

In \cref{Sec:3}, we find the exact, most-general, spherically-symmetric static solution of Einstein's field equations, sourced by an anisotropic fluid, and we outline the basic requirements needed to avoid the central singularity. We then focus on a subclass of such models by choosing a particular equation of state and analyze the null and strong energy conditions. 

In \cref{sect:general}, we select  the general class of regular black-hole   solutions by imposing a  set of minimal constraints, namely  dS behavior in the interior, asymptotically flatness at infinity  and the presence of horizons. We also study the general thermodynamic behavior of these models, discussing the first law of thermodynamics and the appearance of the second-order phase transition. Finally, we investigate photon orbits and the QNMs spectrum in the eikonal approximation. 

In \cref{sect:hayward} and \cref{sect:gaussian} the general discussion is applied   and  the results are confirmed  by revisiting two previously-proposed regular black-hole models, the Hayward and the gaussian-core ones, which appear as particular cases of our general class of models. We finally state our conclusions in \cref{Sec:Conclusions}.

\section{Unified description of space-time and matter inside a black hole}
\label{Sec:2}
As anticipated in the introduction, in the present paper we adopt a description of gravitational interactions in terms of an effective multi-scale field theory, characterized by the generations of hierarchically different length-scales. 

This description is natural in the cosmological and galactic context, as gravity and baryonic matter are characterized by: $1)$ the Planck length $\lP$, $2)$ the size of the cosmological horizon $L$, related to the cosmological constant by $L=\Lambda^{-1/2}$, and $3)$ the gravitational radius of a clump of  baryonic matter with mass $M$, $\RS \sim \lP^2 M$. As already mentioned, an intermediate (mesoscopic) IR length scale in the galactic regime 
  \begin{equation}
  \label{f3}
 R_0= \sqrt{\RS L} 
 \end{equation}
is dynamically generated from $\RS$ and  $L$. At this scale, gravity deviates from its Newtonian behavior as it is evident from rotational curves of galaxies. Moreover,$R_0$ can be seen as a scale at which long-range quantum gravity effects become relevant \cite{Verlinde:2016toy, Cadoni:2017evg, Cadoni:2018dnd,Tuveri:2019zor}.  
This scenario allows for an effective description in the GR framework in terms of an anisotropic fluid, which can be seen as a two-fluid model of dark energy and matter \cite{Cadoni:2017evg, Cadoni:2018dnd,Tuveri:2019zor, Cadoni:2020izk, Cadoni:2020jxe, Cadoni:2021zsl}. The resulting space-time is the SdS solution, in which dark energy dominates at very large scale. In this regime, we have a description in terms of the pure dS space-time and a related scale isometry \cite{Cadoni:2006ww}.
When instead clustered matter $M$ is present and becomes non-negligible, the scale invariance of the dS-background is broken, the quantum scale $R_0$ is generated and we have an effective description in terms of the SdS space-time. The latter is characterized by an internal Schwarzschild-like horizon, determined by the baryonic mass $M$, and by an external dS horizon, which, for small $M$, is located at $r=L$.  The short-scale regime, instead, is described by the Schwarzschild solution with a related scale $\RS$, at which the matter  contribution dominates over dark energy.  The geometry is asymptotically dS \footnote{Notice that, in order to make contact with a black hole space-time, we have to use a static parametrization of the dS geometry.}. 

 In the emergent  gravity scenario of Ref.~\cite{Verlinde:2016toy}, these two regimes are assumed to be endowed with a microscopic description in terms of quantum gravity degrees of freedom (DOFs) entangled on short-scales (at $r\sim \RS$) and on cosmological scales ($r\sim L$). Following Refs.~\cite{Verlinde:2016toy, Cadoni:2018dnd},  the short-range entanglement is responsible for the holographic horizon-area scaling of the entropy. The long-range regime is, instead, characterized by  the slow thermalization of IR, long-range interacting, quantum-gravity DOFs. This  IR dynamics is responsible for an extensive, i.e. \textit{volume}-dependent, contribution to the  entropy. As argued in Ref.~\cite{Verlinde:2016toy}, the competition between the area- and volume-laws in the  entropy  generates a mesoscopic scale $R_0$  and an additional gravitational dark force explaining the deviations from the Newtonian dynamics at galactic scales. 
This multiscale description of gravity, with a "fast scale", $\RS$ and and a "slow scale" $L$, is reminiscent of thermodynamic systems characterized by a glass transition \cite{Verlinde:2016toy} \footnote{ At short  time scales, glassy systems have properties which cannot be distinguished from those of crystals: their effective descriptions are identical. However, the former are characterized by a long timescale behavior, which makes them completely different from crystals.}.\\

Following this line of reasoning, one is led by analogy to use a similar  multiscale description  of matter and gravity for the black-hole interior, in particular to solve the singularity problem. We will consider only macroscopic black holes, i.e. black holes whose horizon radius is hierarchically larger than $\lP$. The short-distance behavior in the black-hole interior (near the singularity) is now dominated by the short-scale dynamics of the emergent space-time DOFs. It is natural to assume that, similarly to the cosmological case at large scales,  here the contribution of matter is negligible at short scales, where we have an effective GR description in terms of a pure dS space-time.  This regime is therefore characterized by an ultraviolet (UV) dS length $\hat L$, a related cosmological constant $\hat \Lambda= \hat L^{-2}$  and scale  invariance. This description is fully consistent with the existence of an UV fixed point, predicted by the quantum-gravity asymptotic safety scenario (see, e.g. Refs. \cite{Bonanno:2000ep, Niedermaier:2006wt,Bonanno:2020bil};  for a resolution of the classical singularity in the asymptotic safety scenario, see Ref. \cite{Adeifeoba:2018ydh}; for recent results on scale invariance in the core of black holes, see Ref. \cite{Borissova:2022jqj}).  Moreover, the dS behavior of the space-time at short scales is consistent with volume-law contribution to the entropy.

Introducing baryonic matter $M$ breaks the scale and conformal invariance of the dS space-time  in the black-hole interior. Similarly to the galactic and cosmological regimes \cite{Tuveri:2019zor}, in this case a new quantum scale $\ell$ is generated in terms of $\hat L$ and $\RS$.  Using an argument similar to that of Ref.~\cite{Verlinde:2016toy}, the generation of $\ell$ can be also explained in terms of the competition between the short-range, volume contribution and the area-law-Bekenstein-Hawking contribution to the entropy we have  at large distances from the center, at the Schwarzschild radius $\RS$.

We see therefore that a multiscale description of gravitational interactions can be adopted both to describe black-holes in a cosmological background and the interior of  asymptotically flat black holes. In the two cases, however, the horizon positions are reversed.  In the latter case, the dS horizon is the internal one, whereas the matter-determined horizon is the external one.  For this reason, even if we expect $\ell=f(\hat L,  \RS)$, this relation needs not to be the same as that relating $R_0$, $L$ and $\RS$  in  \cref{f3}. 
Another difference from the cosmological SdS case, is that here we have the possibility of an external description, i.e. a description of an asymptotic observer at $r\to\infty$. The latter sees "quantum" deviations from the Schwarzschild geometry, parametrized by $\ell$.
In this respect, it should be emphasised once again that the relation between the cosmological case, described by \cref{f3}, and the black-hole case, described instead by \cref{scaling}, is that of an analogy. In particular, this prevents one from finding any relation between the cosmological scales $(L, R_0)$ and our scales $(\hat L, \ell)$.   

From this perspective, we have a new phase in the black-hole interior, in which the emergent gravity DOFs and matter should allow for an effective two-fluids description, i.e. an effective description in terms of an anisotropic fluid \cite{Bayin:1985cd}. 
In the next sections, we will construct a general class of GR models describing gravity sourced by an anisotropic fluid, which allows for non-singular black-hole solutions with two event horizons and an internal dS core.   

\section{Spherically symmetric solutions  sourced by anisotropic fluids}
\label{Sec:3}

Our starting   point is GR  sourced by an anisotropic  fluid. The  stress-energy tensor  $T_{\mu\nu}$  appearing in Einstein's equations $G_{\mu\nu}=8\pi G T_{\mu\nu}$ will be  that pertaining  to  an anisotropic fluid. Anisotropic fluids have a long history and have been fruitfully used in several different contexts in gravitational studies, including compact objects, singular and non-singular black hole models, cosmology (for an incomplete list, see, e.g. Refs. \cite{Cadoni:2017evg, Cadoni:2020izk, Cadoni:2020jxe, Cadoni:2021zsl,cosenza1981some, Bayin:1985cd, DeBenedictis:2005vp, Hayward:2005gi, Nicolini:2005vd, Chirenti:2007mk, Chan:2011ayt, MartinMoruno:2011rm, Aluri:2012re, Culetu:2013fsa, Harko:2013wsa,Raposo:2018rjn, Beltracchi:2018ait, Simpson:2019mud, Kumar:2021oxa, Musco:2021sva, Franzin:2021vnj}). 

We consider static, spherically-symmetric solutions of the theory, whose metric part can be written in the form
\begin{equation}
ds^2 =-e^{\nu(r)}dt^2 + e^{\lambda(r)} dr^2+r^2 d\Omega^2; \hspace{0.5 cm} d\Omega^2 = d\theta^2 + \sin^2 \theta \ d\phi^2.
\label{metrica}
\end{equation}
where $\nu(r)$ and $\lambda(r)$ are metric functions, depending on the radial coordinate $r$ only.

The stress-energy tensor describing the anisotropic fluid can be written as \cite{cosenza1981some}
\begin{equation}
T_{\mu\nu} = \left(\epsilon + p_{\perp} \right)u_{\mu}u_{\nu} + p_{\perp} \ g_{\mu\nu} - \left(p_{\perp}-p_{\parallel} \right)w_{\mu}w_{\nu},
\label{TensoreEI}
\end{equation}
where $\epsilon(r)$, $p_{\parallel}(r)$ and $p_\perp(r)$ are the energy density and the radial and tangential pressure components, respectively, while $u_\mu$ and $w_\mu$ are 4-vectors satisfying the normalization conditions $g^{\mu\nu} u_{\mu}u_{\nu} = -1$, $g^{\mu\nu} w_{\mu}w_{\nu} = 1$ and $u^{\mu}w_{\mu} = 0$.

The independent Einstein's field and stress-energy tensor conservation equations read (the prime denotes derivation with respect to $r$)
\begin{subequations}
\begin{align}
&\frac{1-e^{-\lambda}+re^{-\lambda}\lambda'}{r^2}=8\pi G \epsilon; \label{Einst00}\\
&\frac{e^{-\lambda}-1+re^{-\lambda}\nu'}{r^2} = 8\pi G p_\parallel; \label{grrfieldequation}\\
&p'_\parallel + \frac{\nu'}{2}\left(\epsilon + p_\parallel \right)+\frac{2}{r}\left(p_\parallel-p_\perp \right)=0.\label{TOVeq}
\end{align}
\end{subequations}

Integration of the first equation yields
\begin{equation}\label{g00}
    e^{-\lambda(r)} = 1-\frac{8\pi G}{r} \int \epsilon \ r^2 dr \equiv 1-\frac{2G m(r)}{r}
\end{equation}
where $m(r)$ is the Misner-Sharp (MS) mass
\begin{equation}\label{MSmass}
    m(r) \equiv 4\pi \int_0^r d\tilde r \, \tilde r^2 \,  \epsilon(\tilde r).
\end{equation}

Using \cref{MSmass}, \cref{grrfieldequation} can be recast in the more useful form
\begin{equation}\label{grr}
   \frac{\nu'}{2}=\frac{4\pi G p_\parallel r^3 + Gm}{r\left(r-2Gm \right)}.
\end{equation}

The system \eqref{Einst00}-\eqref{TOVeq} is not closed. In order to determine the solution unambigously, we must  support \cref{Einst00,TOVeq} with two further equations. The simplest and physically natural way to close the dynamical system is to provide: $1)$  a barotropic equation of state (EoS) for the radial pressure $p_\parallel=p_\parallel(\epsilon)$ and $2)$ the matter density profile $\epsilon(r)$. In the following, we will fix the equation of state and the matter density profile by imposing absence of singularities, Schwarzschild behavior at $r\to \infty$   and using the analogy with cosmology discussed in \cref{Sec:2}.

 \subsection{Equation of state and energy conditions}
\label{Sec:4}

The simplest and most natural EoS we can choose is 
\begin{equation}\label{EoSdarkenergy}
    p_\parallel =-\epsilon.
\end{equation}
This choice is physically well-motivated by the  analogy with the cosmological and galactic regime, since it allows both for a dS and SdS (cosmological) phases. It allows for a pure dS behavior near $r=0$,  which implies   the absence of a  singularity in the black-hole interior. Moreover,   the 
EoS also allows for asymptotically-flat solutions at $r\to \infty$, when both $p_\parallel\to 0$ and  $\epsilon\to 0$. One can now easily check that, using   \cref{Einst00} and \cref{grrfieldequation}, the EoS \eqref{EoSdarkenergy} implies $\lambda(r)=-\nu(r)$. In the remainder of the paper, we will adopt the following parametrization of the metric functions $e^{\nu} =  e^{-\lambda}=A(r)$.  

\Cref{grr} can be readily integrated, using \cref{MSmass,EoSdarkenergy}, and yields 
\begin{equation}\label{enu}
    A(r)= 1-\frac{2G m(r)}{r}.
\end{equation}
Finally, using \cref{TOVeq,MSmass,EoSdarkenergy}, we can express the fluid anisotropy $p_\perp - p_\parallel$ as a function of the MS mass as follows
\begin{equation}\label{AnisotropyMSmass}
\frac{p_\perp - p_\parallel}{r} = \frac{1}{4\pi r^3} \left(m'-\frac{r m''}{2} \right).
\end{equation}

It  is useful to write down  explicitly  the energy conditions for the specific case in which the EoS \eqref{EoSdarkenergy} holds. 
\subsubsection*{Null energy condition (NEC)}
In order this condition to be satisfied, we have to require that both $\epsilon + p_\parallel \geq 0$, $\epsilon+p_\perp \geq 0$  hold globally \cite{Hawking:1973uf}. The first is trivially satisfied due to \cref{EoSdarkenergy}, while the second one reduces to
\begin{equation}\label{NullEnergyCondGeneral}
\epsilon'(r) \leq 0,
\end{equation}
upon using \cref{MSmass,AnisotropyMSmass}.
\subsubsection*{Strong energy condition (SEC)}
In this case, we have to require $\epsilon + p_\parallel + 2p_\perp \geq 0$ to hold globally \cite{Hawking:1973uf}. Together with \cref{MSmass,AnisotropyMSmass}, this requirement reduces to
\begin{equation}\label{StrongEnergyCondGeneral}
2 r \epsilon(r) + r^2 \epsilon'(r)\leq 0.
\end{equation}

\subsection{Absence of singularity and  behavior near $r=0$}
\label{Subsec:ConditionsNoSingularity}

In order to avoid the presence of a central singularity at $r=0$  we first impose a set of minimal, very general requirements on the form of the metric functions and  on the density and pressure profiles:
\begin{itemize}
    \item Regularity of  $e^{-\lambda}$  in $r=0$ together  with  \cref{g00,enu} require $m(r) \to 0$ sufficiently fast for $r \to 0$;
    \item 
    Regularity of $p_\parallel$  in $r=0$  together  with \cref{grr} require $r \nu' \to 0$ sufficiently fast for $r \to 0$;
     \item We also require both $p_\parallel$ and $p'_\parallel$ to be smooth in $r=0$. From \cref{TOVeq}, this  implies the regularity condition for the tangential pressure component     
\begin{equation}
        \lim_{r\to 0} \frac{p_\perp-p_\parallel}{r} =0.
        \label{pressurecond}
    \end{equation}
\end{itemize}

Equation \eqref{pressurecond}, together with \cref{AnisotropyMSmass}, implies the following behavior near $r=0$ for the mass function $m(r)$
\begin{equation}\label{massconstraint}
m(r) \sim m_1 + \frac{r^3}{2\ell_\text{P}^2\hat L^2} + \mathcal{O}(r^4) \quad   \text{for} \quad  r\to 0
\end{equation}
with $m_1$ and $\hat L $ some integration constants. Absence of  curvature singularity  for the  metric in $r=0$ requires  $m_1 =0$. The other term, proportional to $r^3$, instead, gives a local dS  solution with  a dS length $\hat L$
\begin{equation}\label{metricnearzerogeneral}
A(r)\sim 1-\frac{ r^2}{\hat L^2}+\mathcal{O}(r^3)\quad   \text{for} \quad  r\to 0. 
\end{equation}
This dS description of the solution core is fully consistent with both the EoS \eqref{EoSdarkenergy} and the analogy with the SdS solution in cosmology (see, e.g. Ref. \cite{Simpson:2019mud} for a regular model with a Minkowski core).

\subsection{Asymptotic flatness and behavior at $r\to\infty$ }
\label{Subsec:ConditionsAtinfinity}

In the asymptotic ($r\to \infty$) region, our solution must be indistinguishable from the Schwarzschild solution. This implies  the space-time to be asymptotically flat, with a subleading Schwarzschild term in the metric function $A(r)$. Moreover, the two pressure components $p_\parallel$ and $p_\perp$ have to vanish in the limit $r \to \infty$. \Cref{AnisotropyMSmass} implies that the minimal condition to have $p_\perp \to 0$ is
\begin{equation}
m = \mathcal{C}_0 r^2+ \mathcal{C}_2 r + M +{\cal O}(\frac{1}{r}),
\end{equation}
with $\mathcal{C}_0,\,\mathcal{C}_2$ and $M$ integration constants.   Asymptotic flatness and absence of  conical defects require $ \mathcal{C}_0=\mathcal{C}_2 =0$, whereas  $M$ becomes the Arnowitt-Deser-Misner (ADM) mass of the solution measured by the asymptotic observer, 

\begin{equation}\label{bc1}
m =  M +{\cal O}\left(\frac{1}{r}\right) \quad \text{for} \quad  r\to \infty.
\end{equation}
We  will also assume that  the function  $m(r)$ interpolates smoothly between the near $r=0$ dS behavior and flat space-time at $r\to \infty$.

\section{A general class of non-singular quantum black-hole models}
\label{sect:general}

Absence of a central singularity and the requirement of a Schwarzschild asymptotic behavior strongly constrain, through \cref{metricnearzerogeneral,bc1}, the local form of the mass function $m(r)$ (or, equivalently, of the metric function $A(r)$) near $r=0$ and $r\to \infty$. However, the global behavior of  $m(r)$, which interpolates between $r=0$ and $r\to\infty$, remains extremely weakly constrained. In this section, we will use the analogy with the  cosmological, SdS case to further constrain its global form. We will get, as output, a general class of models, which can be used to give an effective description of quantum black holes.

Using the analogy with the cosmological case, we assume that $m(r)$ depends only on the two parameters $\hat L$ and $\RS$, which characterize its local behavior near $r=0$ and near $r\to\infty$. This implies that the quantum IR scale $\ell$ can be written as a function of $\hat L$ and $\RS$ only,  $\ell= \ell(\hat L, \RS)$.  The explicit relation between these three scales can be found  using a simple argument. As our models interpolate between the scale-invariant dS behavior in the core and that of clustered matter, which gives the Schwarzschild solution at large radii, there will be a scale $\ell$ at which these two effects balance out. By exploiting  the same arguments used in Ref. \cite{Tuveri:2019zor}, we expect this scale to correspond to the one at which the Compton length associated to a test particle, with mass $m$, in the dS potential $V_{\text{dS}}=r^2/\hat L^2$, is of the same order of the Compton length of the same test particle in the Schwarzschild potential generated by the surrounding mass, i.e. $V_{\text{Sch}}= \RS/r$. The former, curing the singularity at the center, reads $\lambda_\text{c, dS} \sim \hbar/(|V_{\text{dS}}| m) = \hbar \hat L^2/r^2 m$, while the latter, instead, responsible for quantum correction at the horizon scale, is $\lambda_\text{c, Sch} =  \hbar r/\RS m$.  We thus have 
\begin{equation}
\label{f4}
\frac{\lambda_\text{c, dS} }{\lambda_\text{c, Sch}}\sim \mathcal{O}(1) \Rightarrow r\sim \ell \sim \RS^{1/3} \hat L^{2/3},
\end{equation} 
which  is a scaling relation of the form given in \cref{scaling}, with $a=1/3,\, b=2/3$.
We will check this general results in two specific models in \cref{sect:hayward} and \cref{sect:gaussian}.

The presence of a new IR quantum scale implies that we have two complementary descriptions of the quantum black hole. A black-hole-interior description, based on the parameters $\hat L$ and $M$, and a black-hole-exterior one, based instead on $\ell$ and $M$. The second case corresponds to the classical description characterized by the classical hair $M$ and by a quantum deformation parameter, i.e. the quantum hair $\ell$.

A second requirement on the form of the function $m(r)$ comes from the analogy with the cosmological SdS space-time case. The space-time must allow for two horizons, an internal one at $r=r_-$  -- a dS-like horizon -- and an external one at $r=r_+$ -- a Schwarzschild-like horizon. Depending on the value of the parameter $\ell$ (or, equivalently, of the parameter $\hat L$), we will have three cases: $1)$ a black hole with two horizons, $2)$ the two horizons merge and the black hole becomes extremal, and $3)$ a horizonless compact object.

We can easily estimate the relation between the parameters in the extremal case $2)$ using a very simple argument. For the internal observer, which describes its space-time as  dS, the energy density is constant and given by $\epsilon\sim \frac{1}{\hat L^2 \lP^2}$.  The total energy inside a sphere of radius $r$ is $E(r)\sim  \frac{r^3}{\hat L^2 \lP^2}$. For the extremal black hole, we must have $r\sim\hat L$ and the total energy inside the sphere must match the black-hole mass $M$ seen by the outside observer: $E(L) \sim  \frac{\hat L^3}{\hat L^2 \lP^2}=M$. This equation, together with \cref{f4}, implies that the extremal black hole must be characterized by
 \begin{equation}
  \label{f5}
 \ell \sim \RS \sim\hat L. 
 \end{equation}
 We find therefore that $\ell$  has both a quantum origin and should be of the same order of magnitude the Schwarzschild scale $\RS$. 

For $ \ell \lesssim  \RS$,  the presence of the dS core and asymptotic flatness force the metric to have an even number of horizons. In the following, we will limit ourselves to the case of only two horizons  (see below).
In the limit  $\ell\to 0$, corresponding  to  $\hat L\to 0$, the inner dS horizon is pushed towards  $r=0$ and disappears: a singularity is generated in the center. The outer horizon, on the other hand, becomes the classical Schwarzschild one. This case corresponds to the classical limit of our model, in which the usual asymptotically-flat Schwarzschild solution is recovered. In this regime, quantum effects can be  neglected. In view of \cref{bc1}, the simplest way to recover the Schwarzschild solution in the $\ell\to 0$ limit is to assume that   
\begin{equation}
  \label{f9}
 m(r)=m\left(\frac{r}{\ell}, \, M\right).
 \end{equation}

Conversely, in the $\ell \to \infty$ limit, corresponding also to  $\hat L\to \infty$,  the outer horizon disappears and the space-time becomes dS.  Notice that also $M\to \infty$ in this case, consistently with the fact  that we have  a constant energy density. This is the cosmological regime of emergent gravity, in which dark energy in the form of the cosmological constant $L^{-2}$ fully dominates \cite{Verlinde:2016toy}. Our description in terms of a quantum black hole with a dS interior sourced by an anisotropic fluid breaks down in this limit. An effective description of gravitational interactions in terms of GR sourced by an anisotropic fluid is still valid. It can be used to describe galactic dynamics and the generation of the IR length of galactic size \eqref{f3}, giving rise to interesting effects, like the emergence of a dark force at galactic level \cite{Cadoni:2017evg, Cadoni:2018dnd, Tuveri:2019zor}. 
Finally, for $\ell \gtrsim \RS$,  we have a solution with no horizons, which can be thought of as an horizonless star. 

Let  us now write down the most general  form  of the metric  function satisfying   the  conditions discussed   above and in sections  \ref{Subsec:ConditionsNoSingularity} and \ref{Subsec:ConditionsAtinfinity}. We can parametrize the metric function $A(r)$ in terms of a smooth function $F$ as follows:
\begin{equation}
  \label{f10}
A(r)= 1- \frac{\RS}{\ell}F\left(\frac{r}{\ell}\right)\equiv 1-\alpha F(y)
 \end{equation}
where we have defined  the dimensionless coupling $\alpha\equiv \RS/\ell$ and  radial coordinate  $y\equiv  r/\ell$. Furthermore, the function $F$ must satisfy the following conditions: 

\begin{flalign}\label{f12ab}
\quad \, 1. \quad \quad\quad \quad \quad\quad \quad \quad\quad \quad \quad\quad \quad\quad\quad \quad\quad F(y)\sim  \frac{1}{y} \, \quad \text{for}  \  \, y\to \infty;&&
\end{flalign}

\begin{flalign}\label{hhhh}
\quad \,2. \quad \quad\quad \quad \quad\quad \quad \quad\quad \quad \quad\quad \quad\quad\quad \quad \quad F(y)\sim y^2 \, \quad \text{for} \  \, y\to 0;&&
\end{flalign}

\begin{enumerate}
\item [3.] The equation $ 1-\alpha F(y)=0$ admits at most two real positive roots  $y_+$, $y_-$. Moreover, parameter regions discriminated by $\alpha$ are present in which the equation  allows for two  distinct, a single one or no real positive  roots.
\end{enumerate}
When we have two roots $y_+$ corresponds to an event horizon while the inner horizon, given by $y_-$, is generally a Cauchy horizon. The presence of Cauchy horizons raised several concerns in the literature regarding the stability and the viability of such regular black-hole solutions with two horizons. Indeed, according to Poisson, Israel and Ori \cite{Poisson:1989zz, Ori:1991zz}, the Cauchy horizon is typically exponentially unstable under perturbations, an effect known as "mass inflation". In standard regular-black-holes approaches with $\ell \sim \lP$, this instability develops in a time of order of the Planck time, which is a much shorter timescale than the evaporation time, challenging therefore the ability of these models to give  a complete description of the whole evaporation process  \cite{Markovic:1994gy,Carballo-Rubio:2018pmi,Carballo-Rubio:2021bpr}. However, it has been shown recently \cite{Bonanno:2020fgp} that a detailed  analysis is needed in the non-singular black-hole case and that the mass instability does not occur in some specific regular models, like those of Refs. \cite{Hayward:2005gi, Bonanno:2000ep} \footnote{We point the reader to Refs. \cite{Giugno:2017xtl, Casadio:2015qaq,Casadio:2022ndh} for an alternative regular model without the presence of an inner Cauchy horizons in the corpuscular gravity framework.} (but see also Ref. \cite{Barcelo:2022gii} for very recent results).

Apart  from the three conditions above  on the form of the function $F(y)$, we also introduce a constraint on the form of its derivative
\begin{enumerate}
\item[4.] In the region $y \ge y_+$,  $F'(y)/F$ has only one local  extrema (a maximum).
\end{enumerate}
The latter  represents  a statement on the behavior of the black hole temperature and is needed to have the simplest thermodynamic phase portrait (see \cref{sect:bht}). 

 Finally, requiring  the regular quantum-corrected metric to be an exact solution of some effective field equations derived from an action principle could place further constraints on the  function  $F$. These constraints are analyzed in Ref.~\cite{Knorr:2022kqp}. The analysis is mainly concerned with possible quantum corrections to the Schwarzschild metric, assumed to be polynomial in $1/r$ at asymptotic infinity (even if the main results of the paper seem to hold even if the latter assumption is relaxed). These corrections are derived as asymptotic solutions of effective field equations derived from an Einstein-Hilbert action corrected by additional higher-order terms in the curvature. What is found is that, unless either fine-tuning is assumed or strong infrared non-localities in the gravitational action are taken into account, algebraic forms of $F$, like those of Refs. \cite{Simpson:2019mud,Hayward:2005gi}, for instance, are incompatible with a principle of least action. Therefore, their feasibility as "quantum-deformed" black holes may be  questionable, at least if one requires these solutions to be derived from an Einstein-Hilbert action with higher-order terms in the curvature.

Condition $3$ implies the existence of critical values $\alpha_c$ and $y_c$ for $\alpha$ and $y$ respectively, labeling the extremal case, when the two black-hole horizons merge. These critical values are determined by requiring the metric function to have a double-degenerate zero, i.e by the system of equations,
\begin{equation}
  \label{f11}
 1-\alpha F(y)=0,\quad F'(y)=0.
 \end{equation}

Notice that a simple principle of naturalness implies $\alpha_c$ and $y_c$ being of order $1$, so that the critical values for $\ell$ and $r$ are of order $\RS$. Therefore, $\alpha_c$ allows us to distinguish three  regimes for our model:
\begin{itemize}
\item[$a)$] $\alpha\gg\alpha_c$, corresponding  to $\ell\ll \RS$ (i.e $\ell\sim \lP$),  which describes the Schwarzschild black hole with small quantum corrections \cite{Modesto:2010uh, Nicolini:2005vd, Spallucci:2011rn,Modesto:2004xx, Modesto:2008im, Nicolini:2008aj, Hossenfelder:2009fc,Sprenger:2012uc,Bambi:2013caa,Frolov:2014jva,Casadio:2014pia,Binetti:2022xdi};
\item[$b)$] $\alpha\sim \alpha_c$, corresponding to $\ell \sim \RS$. In this case, $\ell$ parametrizes  quantum gravity effects at horizon scales \cite{Mathur:2005zp, Mathur:2019dhf,Mathur:2020ely,Verlinde:2016toy,Dvali:2011aa, Dvali:2013eja, Dvali:2020wqi, Casadio:2015lis, Casadio:2016zpl, Cadoni:2020mgb,Maggiore:2007nq,Cadoni:2021jer, Cadoni:2021qfn};
\item[$c)$] $\alpha<\alpha_c$, corresponding to $\ell > \RS$, which describes horizonless stars. We will not consider this case in the following.
\end{itemize}

\begin{figure}
\centering
\includegraphics[width= 9 cm, height = 9 cm,keepaspectratio]{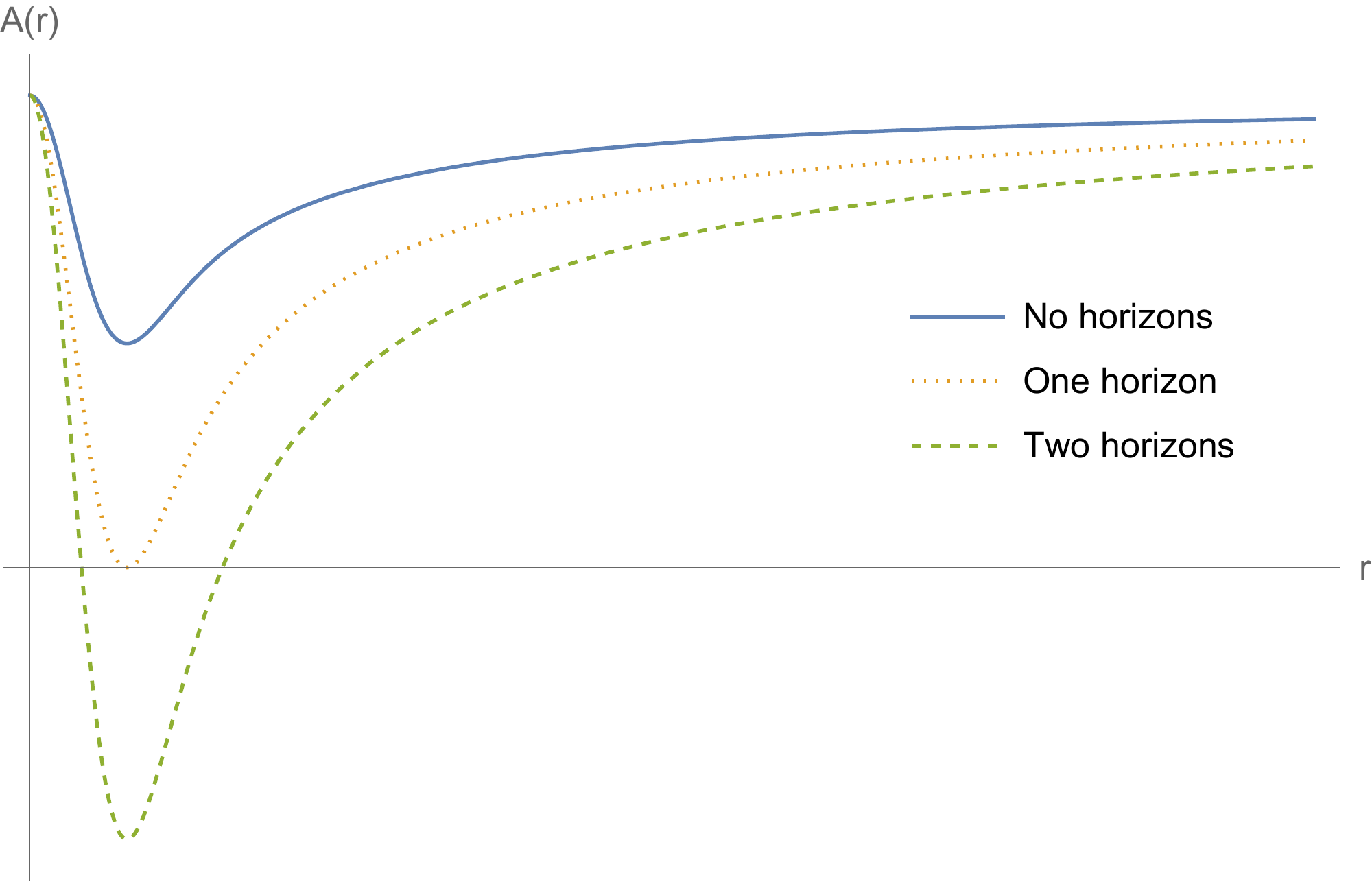}
\caption{Qualitative behavior of the metric function $A(r)$ as a function of the radial coordinate and for different values of the parameter $\alpha$. We can either have solutions with no horizons for $\alpha<\alpha_c$ (blue solid curve), extremal solutions with a single horizon for $\alpha=\alpha_c$ (orange dotted curve) or solutions with two horizons for $\alpha>\alpha_c$ (green dashed curve).}
	\label{QualitativeMetric}
\end{figure}
A qualitative behavior of the metric function $A(r)$    in the three cases is plotted in \Cref{QualitativeMetric}.\\
\subsection{Energy conditions}
Given the form  \eqref{f10} for the general metric function  and  taking into account \eqref{f12ab}  and \eqref{hhhh}, we  can 
rewrite  the two energy conditions \eqref{NullEnergyCondGeneral} and \eqref{StrongEnergyCondGeneral} in terms of the function $F$ and discuss their behavior near $r \sim 0$ and for $r \to \infty$. Using \cref{g00,f10} in  \eqref{NullEnergyCondGeneral} and \eqref{StrongEnergyCondGeneral}, we get
\begin{subequations}
\begin{align}
&\epsilon'=-\frac{\alpha F}{4\pi G \ell^3 \, y^3}+\frac{\alpha F''}{8\pi G \ell^3 \, y}\le 0;\\
&2 r \epsilon + r^2 \epsilon' = \frac{\alpha \left(2 F' + y F'' \right)}{8\pi G \ell}\le 0.
\end{align}
\end{subequations}
Near $y \sim 0$,  $\epsilon'=0$ follows, so that  the NEC is satisfied, while $2 r \epsilon + r^2 \epsilon' \sim 0^+$ for $y\to 0$: the SEC is violated somewhere deep in the core of the astrophysical object. This is expected, since this violation is a characteristic of the dS spacetime.

On the other hand, for $y\to \infty$, 
using \cref{f12ab}  we get  $\epsilon'\sim 0$ and $2 r \epsilon + r^2 \epsilon'\sim 0$. Thus,  the NEC and the SEC are satisfied  in the asymptotic region,  as expected  in view of the Schwarzschild asymptotics of the solutions. 

Violation of the SEC in the inner core explains how in our models the usual singularity theorems can be circumvented. Even if the dS behavior near $r=0$ assures the absence of a curvature singularity at $r=0$, in principle it is not sufficient to guarantee the geodesic completeness of the space-time described by the metric function  \eqref{f10}. In \cref{appendix} we explicitly show that in  our models, caustics  of time-like geodesics cannot form,  proving therefore the geodesic completeness of space-times described by \cref{f10}. 

\subsection{Extremal limit}
\label{sec:ExtremalLimitGeneral}

As it is usually the case for standard charged and/or rotating black holes \cite{Giddings:1992kn,Bardeen:1999px,Hartman:2008pb,Kunduri:2013gce}, in the extremal limit and in the near-horizon approximation, the local geometry of our  space-time behaves as AdS$_2 \times$ S$^2$, i.e as the tensor product of a two-dimensional (2D) AdS space-time and a two-sphere, with both the AdS$_2$ length $L_2$ and the radius of S$^2$ of order $\RS$. In fact, in the extremal limit $r_+=r_- =r_c$, the metric function $A(r)$ must have a double zero at $r=r_c$, determined by the solution of \cref{f11}. Expanding it near the horizon, i.e. in power series of $r-r_c$, we get at leading order  
\begin{equation}
  \label{f51}
ds^2=- L_2^{-2}(r-r_c)^2 dt^2+ L_2^{2}(r-r_c)^{-2} dr^2+r_c^2 d\Omega^2,
\end{equation}
where  we have defined $L_2^{-2}=-\frac{1}{2}A''(r_c)$ and  $r_c\sim \RS$ owing to \cref{f5}. Moreover, for purely dimensional reasons, the same equation implies $A(r_c)''\sim \RS^{-2}$, from which $L_2\sim \RS$ follows. A translation of the radial coordinate $r\to r+r_c$ brings the metric into the form  
 \begin{equation}
  \label{f6}
 ds^2=- \frac {r^2}{L_2^{2}} dt^2+ \frac {L_2^{2}}{r^2}dr^2+r_c^2 d\Omega^2,
 \end{equation}
which describes an AdS$_2 \times$ S$^2$ space-time, with the AdS$_2$ metric written in Poincar\'e coordinates. 

As we will see  later in this paper, the extremal solution is a zero-temperature, zero-entropy  solution. The extremal configuration will be then thermodynamically preferred. Even if a solution with two horizons could  result from astrophysical collapse of a compact object, it will decay in a time much shorter than the Hawking evaporation time into the extremal configuration.

Let us conclude this subsection by noting that the extremal solution is stabilized by a particular profile for the energy density $\epsilon$ and for the pressures $p_\parallel$   and   $p_\perp$. In the near-horizon approximation, when the metric takes the simple AdS$_2 \times$S$^2$ form, we expect them to be constant and to satisfy  a simple EoS. In fact, combining \cref{TOVeq,g00,f10}, the extremality conditions \eqref{f11} and the EoS \eqref{EoSdarkenergy}, we find that the leading terms for the near-horizon energy density and pressures are

\begin{subequations}\label{epsilonpressuresnearhorgeneral}
\begin{align}
\epsilon^{\text{(nh)}}&= \frac{1}{8\pi G r_c^2}, &&p_\parallel^{\text{(nh)}}=-\epsilon^{\text{(nh)}},\\
\epsilon^{\text{(nh)}}_{\text{(AdS)}}&= -\frac{1}{8\pi G L_2^2}, &&p^{(\text{nh})}_\perp =-\epsilon^{\text{(nh)}}_{\text{(AdS)}},
\end{align}
\end{subequations}

where $\epsilon^{\text{(nh)}}_{\text{(AdS)}}$ is the (negative) constant energy density sourcing  AdS$_2$. 
It is quite interesting to note that both the radial and perpendicular components of the pressure satisfy  the simple equation of state $p=-\epsilon$.  The (positive) energy density associated to the two-sphere acts as a source for the (negative) radial pressure, whereas  the (negative) energy density associated to AdS$_2$ acts as source for the (positive) perpendicular pressure. Thus, the stabilization of the AdS$_2 \times$S$^2$ near-horizon, extremal solution is achieved in a rather non-trivial way.

\subsection {Black hole thermodynamics}
\label{sect:bht}
From the metric function \eqref{f10}, using standard formulae, we can compute both the Hawking temperature $T_\text{H}$ and the black-hole mass for the quantum corrected black hole, as functions of the outer horizon radius $r_+\equiv \rH$ and of  quantum deformation parameter $\ell$: 
\begin{equation}\label{THQBHgeneral}
T_\text{H}(\rH,\ell) = \frac{1}{4\pi} \frac{d A(r)}{dr}\biggr|_{r = \rH}=-  \frac{\alpha}{4\pi \ell}F'\biggr|_{y =y_\text{H}}, \qquad M (\rH,\ell)= 
\frac{\ell}{2G}F^{-1}(y_\text{H}).
\end{equation}
An important point is that $\ell$ has to be considered as a quantum-deformation parameter which, contrary to $M$, is not associated with conserved charges. This makes our quantum black-hole solution drastically different from other two-parameter classes of solutions, like e.g. the charged RN solution, for which \textit{both} parameters are associated with thermodynamic potentials.

Owing to this feature, we look for a first law of thermodynamics of the form $dM = T_\text{H} dS$, where $S$ is the black hole entropy. One can easily check that the area law, i.e an entropy equal to a quarter of the area of the outer event horizon (in Planck units), cannot be valid for our class of black-hole models. In fact, using $S_A = \mathcal{A}_\text{H}/4G= \pi \rH^2/G$, we get $dM-T_\text{H}dS_A=-(1/2G)(dF/dy_\text{H}) F^{-1}\left(F^{-1}-\rH/\ell\right)d\rH$. This tells us that, once the area law is assumed, the first principle is satisfied only for $F(y)= \ell/\rH$, i.e only for the Schwarzschild black hole.

Let us now look for a new  definition  of  the black hole entropy $S$, generalizing  the area law, such that the first  principle is satisfied. This generalized entropy can be found by noticing that \cref{THQBHgeneral} implies the validity of the following relation

\begin{equation}\label{gg2}
dM=4\pi  M T_\text{H}d\rH.
\end{equation}
By  defining the  black hole entropy as
\begin{equation}\label{gg3}
S=4\pi  \int M d\rH,
\end{equation}
we see that the first  principle $dM=TdS$ is satisfied. Moreover, the entropy \eqref{gg3} correctly reproduces the area law in the Schwarzschild case, $M= \rH/2G$. Equation \eqref{gg3}  defines  the entropy of the hole up to an integration constant, which can be fixed by requiring  the entropy area law   to be  recovered in the limit $\ell\ll \RS$, i.e. in the classical limit of our quantum model. This leads to
\begin{equation}\label{Entropydiffgeneral}
S(\rH)=4\pi\int^{\rH}_{r_{m}}  M(\rH') d\rH', 
\end{equation}
where $r_m$ is the minimum value of the horizon radius. In the limit  $\ell\ll \RS$, we have  $M(\rH)= \rH/2G$ and $r_{m}=0$, so that \cref{Entropydiffgeneral} gives the area law $S= \pi  \rH^2/G$.

For a generic quantum deformed black hole, $r_{m}$ is given by the radius  $r_c$ of the extremal black hole. This implies in particular that the extremal black hole has zero entropy, i.e $S(r_c)=0$. The extremal limit for our quantum deformed black hole is therefore a state of non-vanishing mass, but with zero temperature and entropy. Again, this behavior is drastically different from that of usual extremal black holes, for which  the extremal configuration is a state with $T=0$, but with $S\neq 0$.

Both for large black-hole radii $\rH\to \infty$ and in the extremal limit, the temperature goes to zero. This can be easily checked using \cref{THQBHgeneral} together with  \cref{hhhh} and   \cref{f11}.  Smoothness of the function $F(y)$ then implies that the function $T_\text{H}(\rH)$ has at least one local maximum in the range $[r_c$, $\infty)$. In order to avoid an oscillating behavior of $T_\text{H}(\rH)$, we have restricted ourselves to the simplest case by imposing condition $4$ on the form of the function $F$ (see the beginning of \cref{sect:general}). 

\begin{figure}
\centering
\includegraphics[width= 9 cm, height = 9 cm,keepaspectratio]{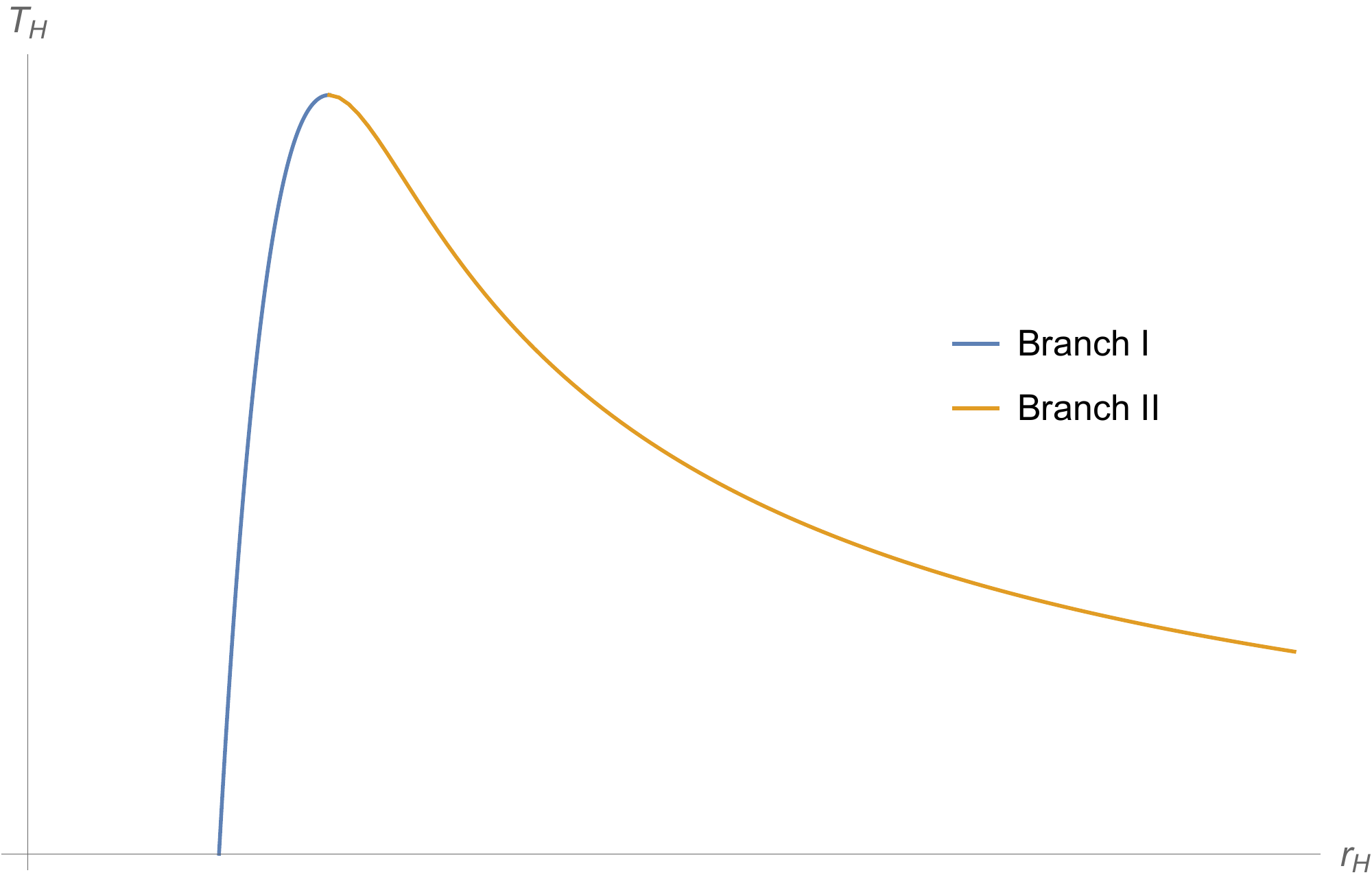}
\caption{Typical  qualitative  behavior of the black hole temperature $T_\text{H}$ as a function of the black hole radius $\rH$. We explicitly show the presence of the two  branches.}
	\label{Temp1}
\end{figure}

The temperature starts form zero in the extremal limit, then it reaches a maximum $T_{\text{H, max}}$ for $r_{\text{H, max}}$ and then goes down to zero again for large values of $r_{\text{H}}/\ell$. This  implies that $T_{\text{H}}$ is always bounded, $0\le T_{\text{H}}\le T_{\text{H, max}}$. Only when we take the limit $\ell\to 0$ first, to recover the Schwarzschild black hole, can the temperature become arbitrarily large by taking small black holes, $r_{\text{H}}\to 0$. Notice that a non-vanishing quantum deformation parameter, $\ell\neq 0$ solves, as expected, the singular thermodynamical behavior $T_\text{H}\to \infty$ of the Schwarzschild black hole for $r_{\text{H}}\to 0$. The typical qualitative behavior of the temperature is shown in \Cref{Temp1}.

In order to study in detail the thermodynamic behavior of the black hole near extremality, we expand $T_\text{H}$ and $M$ near $r_c$. At leading order, we get $T_\text{H}=\gamma (\rH-r_c)$, whereas $M= M_c+ \beta (r_\text{H}-r_c)^2$, where $\gamma=dT_\text{H}/d\rH|_{r_c}$, $\beta= (1/2)d^2M/d\rH^2|_{r_c}$ and $M_c= M(r_c)$. Notice that $dM/d\rH$ is always positive and becomes zero at extremality, $dM/d\rH|_{r_c}= 0$. For this reason, the linear term in the expansion of $M$ is absent. The previous expression implies a quadratic scaling  with the temperature of the mass above extremality  
\begin{equation}\label{qscalinggeneral}
 M-M_c\sim \frac{\ell^3}{G}T_\text{H}^2,
\end{equation}
which is fully consistent with the AdS$_2\times$ S$^2$ near-horizon behavior of the extremal limit \cite{Maldacena:1998uz,Almheiri:2014cka,Almheiri:2016fws}. This means that, in the near-extremal limit, the black hole allows for an effective description in terms of a 2D dilaton gravity theory, i.e. Jackiw-Teitelboim (JT) gravity, with the dilaton parametrizing the radius of the 2-sphere S$^2$ \cite{Jackiw:1984je, Teitelboim:1983ux, Grumiller:2002nm}. This in turn implies the possibility of using a  dual 2D conformal field theory to describe the near-horizon regime of the near-extremal black hole. This fact may be relevant for applications  to the information puzzle in black-hole physics \cite{Hawking:1976ra,Page:1993wv,Mathur:2009hf,Kitaev:2017awl,Almheiri:2019qdq,Penington:2019kki,Almheiri:2019hni}.  \\

\subsection{Phase  transition}
The non-monotonic behavior of $T(\rH)$, which is common to a wide class of charged and/or AdS black holes, signalizes a non-trivial thermodynamic phase structure, the presence of two thermodynamic branches and a phase transition at the critical temperature $T_{\text{H, max}}$ (see, e.g. Refs. \cite{Hawking:1982dh,Pavon:1991kh, Witten:1998zw,Chamblin:1999hg,Wu:2000id,Cadoni:2009xm,Kubiznak:2012wp,Rajagopal:2014ewa,Mandal:2016anc,Li:2020nsy}). This simply follows from the fact that we have two possible values of $\rH$ for a given value of $T_\text{H}$. This implies the presence of metastable states and the existence of two branches, $I$ and $II$. Branch $I$ corresponds to small, order $1$ values of the dimensionless black hole radius $y_\text{H}$ (the left hand  region of \Cref{Temp1}).  In this branch, $r_\text{H}$ varies between the extremal value $r_c$ and $r_{\text{H, max}}$. Correspondingly, the parameter $\ell$ takes values between $\ell_{\text{H, max}}$ and $\ell_c  $, both of order $\RS$. Thus, branch $I$ describes quantum black holes whose quantum deformation parameter $\ell$ is of the same order of magnitude of the classical Schwarzschild radius $\RS$. Conversely, branch $II$ corresponds to large values of $y_\text{H}$  (the right hand region of \Cref{Temp1}). Here, $\rH$ can take values much larger than $r_{\text{H, max}}$. This corresponds to small values of the parameter $\ell$. Thus, the far right region of branch $II$ describes classical black holes, with quantum deformation parameter $\ell\ll \RS$.

The phase transition and the stability of the different thermodynamic phases can be investigated by considering the specific heat of the black hole:
\begin{equation}\label{sh}
C= \frac{dM}{dT} =\frac{dM}{d\rH} \left(\frac{dT}{d\rH} \right)^{-1}.
\end{equation}
Being $dM/d\rH$ always positive, the non-monotonic behavior of $T_\text{H}$ implies that  
\begin{itemize}
\item For $r_c< r_\text{H}<r_{\text{H, max}}$, $dT/dr_\text{H}$ is positive and thus $C>0$;
\item For $r_\text{H}>r_{\text{H, max}}$, $dT/dr_\text{H}$ is negative and thus $C<0$; 
\item For $r_\text{H}=r_{\text{H, max}}, \,dT/dr_\text{H}=0$  and thus $C\to \infty$. 
\end{itemize}

\begin{figure}
\centering
\includegraphics[width= 8 cm, height = 8 cm,keepaspectratio]{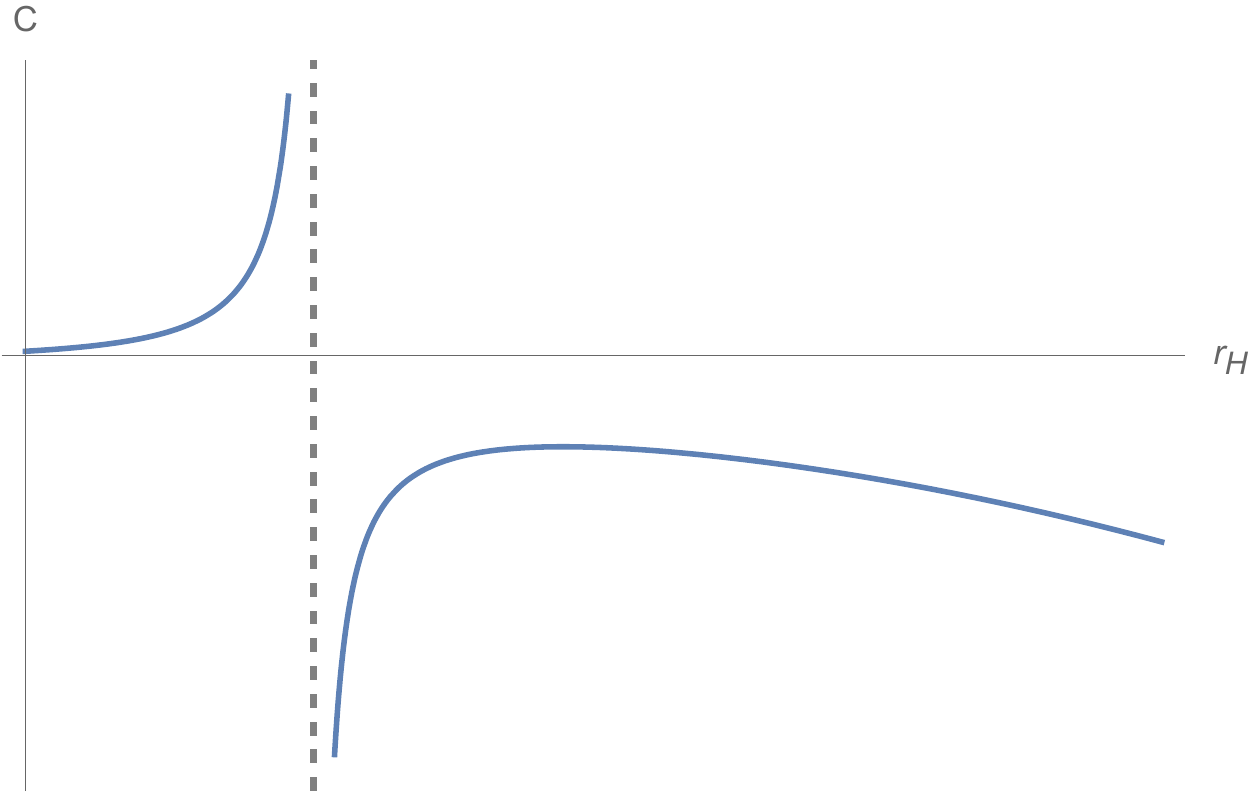}
\caption{Typical qualitative behavior of the black hole specific heat $C$ as a function of $\rH$. The $\rH$-axis starts from the value at which $C$ is zero, i.e. at the extremal value of $\rH$, given by the solution of \cref{f11}. The vertical dashed line, where $C$ diverges, corresponds to the zero of $dT/d\rH$, i.e. the maximum of the temperature.}
	\label{SpeHeat1}
\end{figure}

In \Cref{SpeHeat1} we plot the qualitative behavior of the specific heat $C$ as a function of $\rH$. 

In branch $I$, i.e for values of $\rH$ less than $r_{\text{H, max}}$, the specific heat is  always positive and the black hole exists in equilibrium with its radiation. On the other hand in branch $II$, i.e for values of $\rH$ larger  than $r_{\text{H, max}}$, the specific heat is  always negative and the black hole cannot be considered at equilibrium with its radiation anymore. A second order phase transition occurs at $r_\text{H, max}$. This implies that quantum black holes with small values of the quantum deformation parameter are thermodynamically less favored with respect to black holes with $\ell\sim \RS$. Moreover, the value of the deformation parameter $\ell$ at the maximum, $\ell_{\text{H, max}}$, is very close to the extremal value. This means that black holes close to extremality are thermodynamically preferred, which further confirms the choice $\ell \sim G M$. This also shows that the outcome of the evaporation process will be a \textit{cold} remnant at zero temperature and zero entropy. The latter, in particular, is again an intriguing property, as it could allow to circumvent problems on the viability of remnants as possible solutions on the information paradox \cite{Chen:2014jwq}.

These thermodynamic aspects and the phase picture will be confirmed later on this paper, when we will consider particular cases of our general class of models and we will study their free energy.\\

The non-trivial phase space structure analyzed above is a consequence of the multiscale description of our models, discussed in \cref{Sec:2}. In light of the similarity between this description of gravitational interactions and glassy systems \cite{Verlinde:2016toy}, one could ask whether  our phase transition could be interpreted as a glass transition. Indeed, even the latter is characterized by a divergence in the specific heat at the transition point, but it is not generally classified as a second-order phase transition. To answer this question, one would need first to define some Ehrenfest equations, to describe variations of the specific heat and the derivatives of the volume between the two phases. For a second-order phase transition, both these equations are satisfied, while either both or one of them is violated in glassy systems. In the black-hole case, one can define Ehrenfest-like equations \cite{Banerjee:2010qk} by replacing the volume with the electric charge and/or the angular momentum (if the model is charged and/or rotating) and analyze their variations between the two phases. 

Here, we can only speculate that the phase transition of our "quantum-corrected" black hole could be very similar to a glass transition instead of a second-order one.  In fact, the absence of any thermodynamic potentials related to $\ell$ or other physical observables prevents us from properly define Ehrenfest-like equations, and therefore to perform an analysis similar to the one in Ref.~\cite{Banerjee:2010qk}. Consequently, this does not allow us to assess quantitatively the nature of the phase transition in our models. \\

Summarizing the results obtained so far, the stable configuration of our quantum-black-hole model, realized using an anisotropic fluid, will be represented by an extremal (or near-extremal, if we consider small deviations from extremality) black hole. The black-hole geometry interpolates between a dS  space-time in the black-hole interior (near the singularity), an AdS$_2 \times$ S$^2$ geometry in the near-horizon region, and flat space-time in the asymptotic, $r\to\infty$, region. The dS behavior near $r=0$ solves  the singularity problem. At extremality, the two (dual) quantum scales characterizing the quantum black hole ($\ell$, $\hat L$) have the same order of magnitude of the classical black-hole radius $\RS$.

The scale $\ell$ characterizing quantum effects seen by an external observer is naturally of the order of magnitude of the classical Schwarzschild radius $\RS$ of the black hole. This opens the possibility of having phenomenological quantum signatures, potentially observable in the near future, e.g. through the QNMs spectrum and the geodesic motion of particles near the horizon. Moreover, the near-horizon AdS$_2 \times$ S$^2$ behavior is very promising for tackling the black-hole information puzzle. 


\subsection{Null geodesics and photon orbits}
\label{sect:po}

To compute photon orbits in our class of models, we start from the geodesic equation together with the null-geodesic constraint (dot will refer to derivation with respect to the affine parameter)
\begin{equation}\label{geodesicequation}
\ddot{x}^\mu + \Gamma^\mu_{\nu\lambda} \dot{x}^\nu \dot{x}^\lambda=0, \qquad g_{\mu\nu}\dot{x}^\mu \dot{x}^\nu=0.
\end{equation}

The isometries of the metric (spherical symmetry and invariance under time translations) imply two conservation equations, which by considering geodesics on the plane $\theta = \text{constant}=\pi/2$, take the form
 \begin{equation}\label{g22}
\dot{\phi} = \frac{J}{r^2},\quad \dot{t}=\frac{\mathcal{C}}{|A(r)|}
\end{equation}
where $J$ and $\mathcal{C}$ are  integration constants.

The geodesic equation for the coordinate $r$ can be integrated to give,
\begin{equation}
\dot{r}^2+\frac{J^2}{r^2}A(r)=\mathcal{C}^2,
\end{equation}
which can be rewritten as the   energy  conservation equation: 
\begin{equation}
\frac{1}{2}\dot{r}^2 + V(r) = \frac{\mathcal{C}^2}{2}\equiv \mathcal{E},
\end{equation}
with $\mathcal{E}$ a constant and $V(r)$ the effective potential 
\begin{equation}\label{effpotphotonorbitsgeneral}
V(r) \equiv \frac{J^2}{2 r^2}A(r).
\end{equation}

Since the leading term of metric function for $r\to 0$ is $A(r)\sim 1$, we have $V(r) \to \infty$ in this limit. This behavior is completely different from the Schwarzschild case ($V\to -\infty$  as  $r\to 0$) and is a consequence of the absence of a singularity.   Conversely,  for $r\to \infty$,  $A(r)$ is dominated by the $1/r$ term and $V(r)\to 0$, as in the Schwarzschild case. The shape of the effective potential  $V(r)$ at intermediate distances depends crucially on the values of the parameter $\alpha$. The local extrema of $V$ are the solution of the equation 
\begin{equation}\label{minpot}
r A'(r) -2A(r)=2\alpha F-2  -\frac{\alpha r}{\ell}F'=0.
\end{equation}

Being  $F$ and $F'$  both bounded, it will always exist a mimimum value 
$\alpha_m<\alpha_c$   such that  for $\alpha\le \alpha_m$ the equation has no (real) zeroes and  bounded photon orbits do not exist. For $\alpha>\alpha_m$, the equation allows instead for two (real) zeroes, corresponding to a  local minimum and  a  local maximum for $V(r)$. On the other hand, for  $\alpha_m<\alpha<\alpha_c$, we still have the stable and unstable photon orbits, but there are no horizons, since the solution describes a  star. For  $\alpha=\alpha_c$, i.e in the  extremal configuration, the miminum of the potential coincides with the horizon position. In this case, we will have both the outer unstable photon ring and a stable one, which, however, coincides with the event horizon of the extremal black hole. Finally, for $\alpha>\alpha_c$, the local minimum is inside the event horizon and we have only a bounded unstable photon orbit, similarly to the Schwarzschild case.   

\begin{figure}
\centering
\subfigure[$\alpha> \alpha_c$]{\includegraphics[width=6cm]{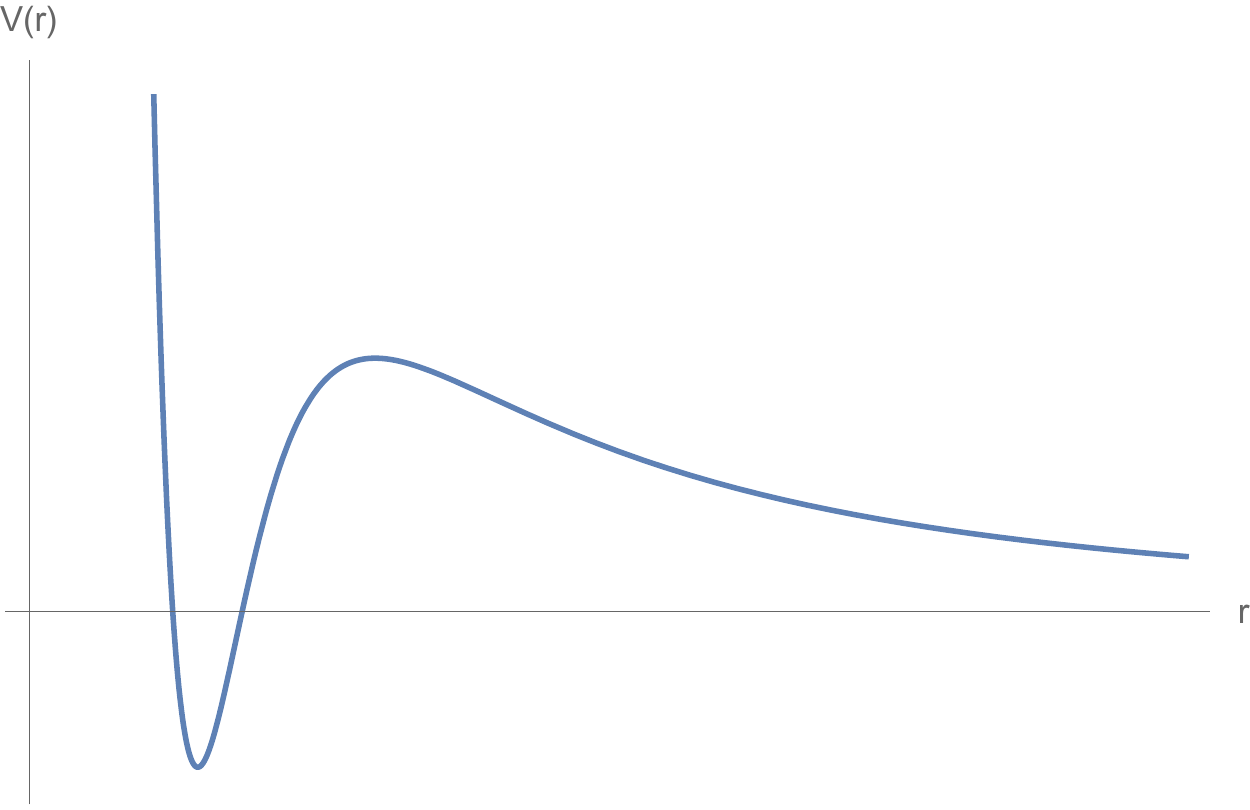}}
\hspace{0.7 cm}
\subfigure[$\alpha = \alpha_c$]{\includegraphics[width=6cm]{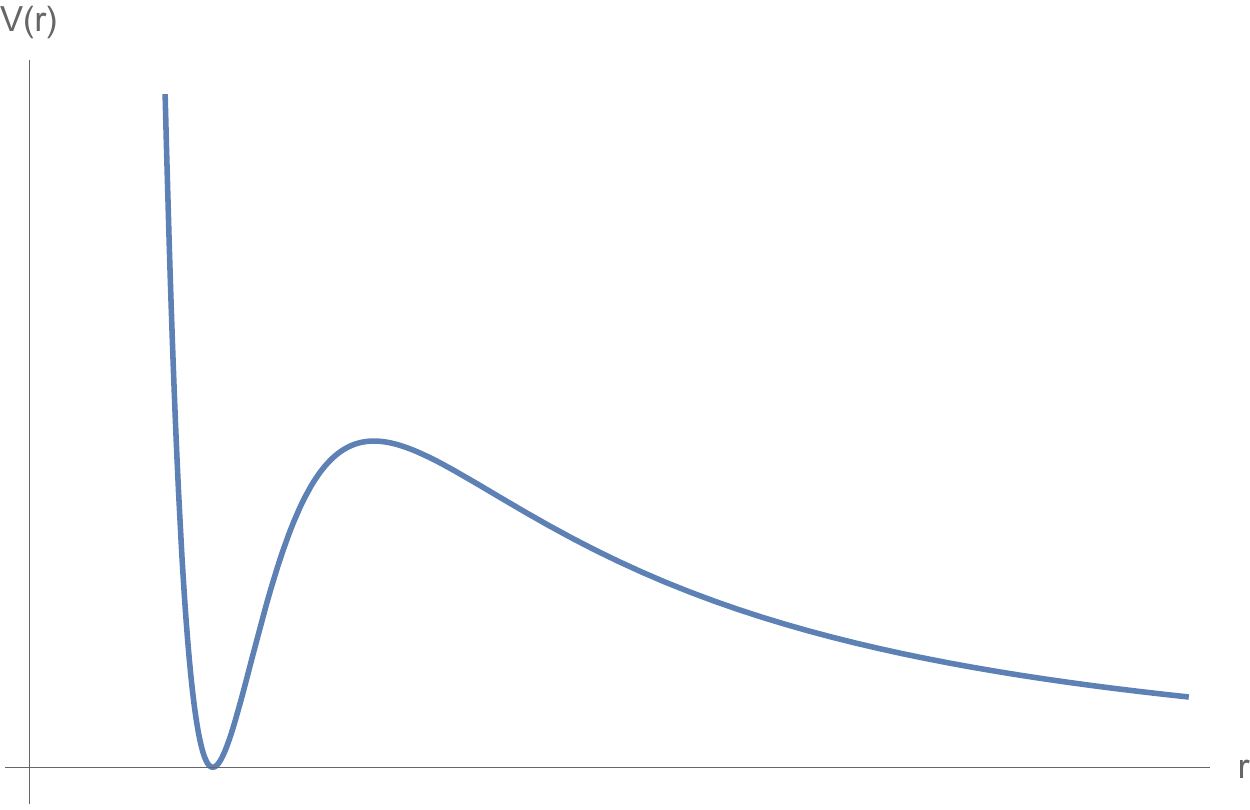}}
\vfill
\centering
\subfigure[$\alpha_m<\alpha<\alpha_c$]{\includegraphics[width=6cm]{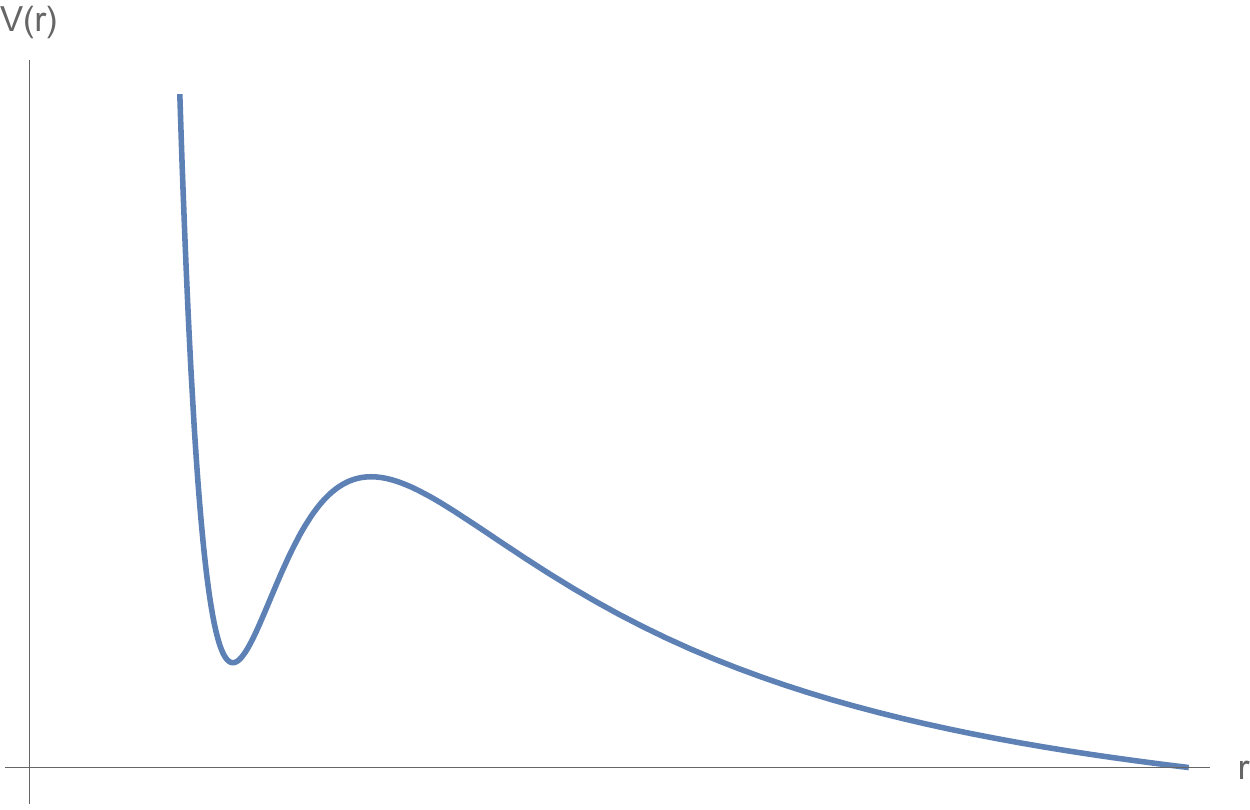}}
\hspace{0.7 cm}
\subfigure[$\alpha< \alpha_m$]{\includegraphics[width=6 cm]{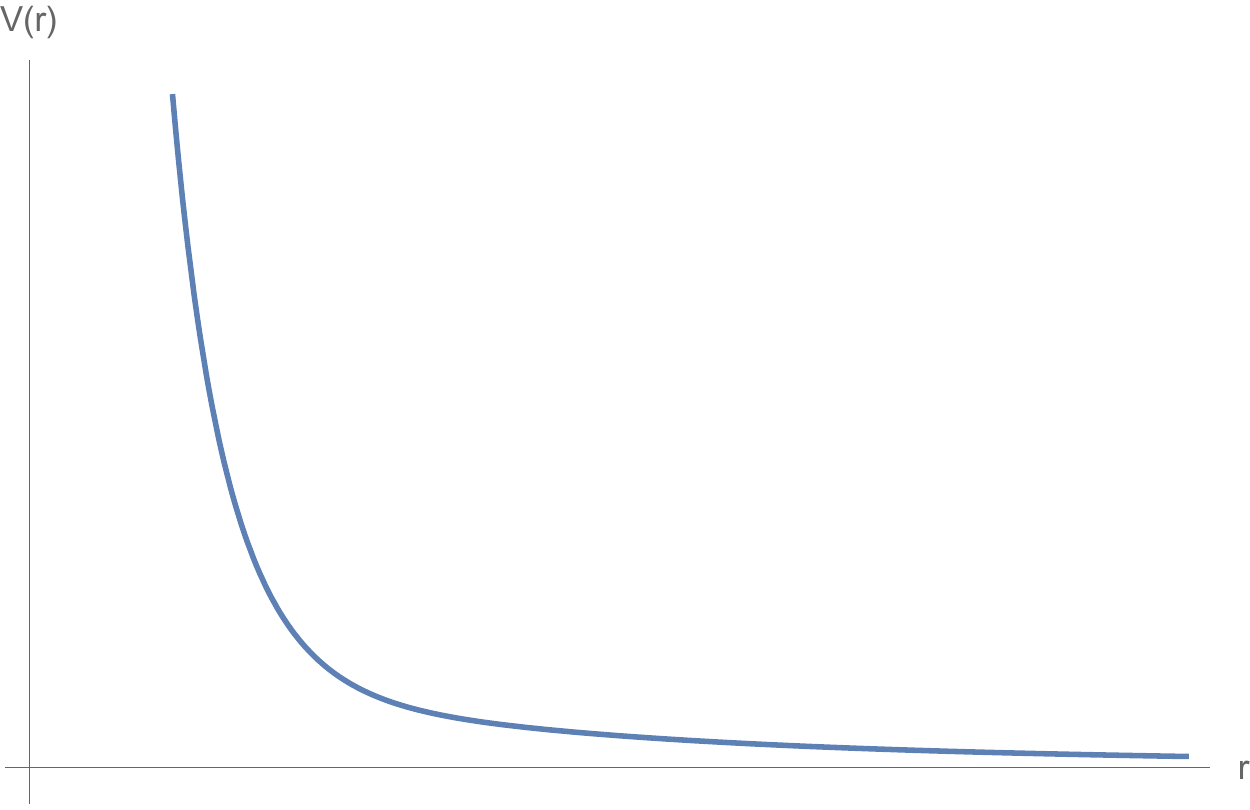}}
\caption{Typical  qualitative  behavior of the effective potential  $V$ for null geodesics as a function of $r$ for $\alpha>\alpha_c$,  $\alpha=\alpha_c$, $\alpha_m<\alpha<\alpha_c$ and   
$\alpha  < \alpha_m$. }
\label{photonOrbgeneral}
\end{figure}

The qualitative behavior of $V(r)$ in the four cases is shown in \Cref{photonOrbgeneral}. 
From the plots we see that there are two main differences with the Schwarzschild case. The first is the possibility of having a complete absence of bounded orbits, in the case of the horizonless solution. The second is the presence of stable orbits for photons in addition to the usual unstable ones. However, in the two-horizon case, the stable orbits are beyond the black-hole horizon and do not play any role, while in the extremal case they coincide with the horizon. Deviations from extremality, however, push this minimum inside the outer horizon and their effects are irrelevant from the observational point of view. 

\subsection{Quasi-Normal Modes spectrum in the eikonal approximation}
\label{sect:qnm}
      
In this section we consider QNMs for scalar perturbations in the fixed background given by our general black-hole solution. We will then use the eikonal approximation to give an analytical estimate of the quasi-normal frequencies. 

The evolution of scalar perturbations $\Psi$, in the fixed gravitational background metric $g^{(B)}_{\mu\nu}$ described by the metric function \eqref{f10}, is determined by the Klein-Gordon equation 
\begin{equation}
\label{KGgeneral1}
\Box \Psi=\frac{1}{\sqrt{-g^{(B)}}}\partial_\mu\left(\sqrt{-g^{(B)}}g_{(B)}^{\mu\nu}\partial_\nu\right)\Psi=0.
\end{equation}
By separating the time,  angular and radial parts of $\Psi$, one gets the radial Schr\"odinger-like equation  for the $r$-dependent part, 
\begin{equation}
\label{KGgeneral1}
\frac{d^2 \psi}{dr_\ast^2}+\left[\omega^2-V_{\text{KG}}(r) \right]\psi=0,
\end{equation}
where the potential is
\begin{equation}
V_\text{KG}(r) = (1-\alpha F) \left[\frac{l(l+1)}{r^2} - \alpha \frac{ F'}{\ell r} \right],
\end{equation}
$l$ is the orbital-angular-momentum number and $\rst$ is the tortoise coordinate  defined by 
\begin{equation}\label{s1a}
\rst=\int \frac{dr}{A(r)}.
\end{equation}
Analytical estimates of the quasi-normal frequencies can be obtained by using an intriguing relation between quasi-normal modes and the parameters chracterizing null geodesics in the eikonal limit $l \gg 1$ first noted in Ref. \cite{Press:1971wr}.
The vibration modes of the black holes, whose energy is gradually being radiated away, are interpreted as photons moving along an unstable null-geodesics and slowly leaking out. This correspondence was more recently analyzed in Refs. \cite{Ferrari:1984zz,Mashhoon:1985cya,Cardoso:2008bp} and shown to agree with WKB results \cite{Churilova:2019jqx,Schutz:1985km}. Specifically, the angular velocity $\Omega_m=\dot{\varphi}/\dot{t}|_{r=r_m}$, computed at the maximum of the geodesic potential $r_m$, determines the real part of the quasi-normal spectrum. Further,  the so-called Lyapunov exponent $\lambda = \sqrt{\frac{V''(r)}{2 \dot{t}^2}}\biggr|_{r=r_m}$ ($V(r)$ is the effective potential for null geodesics \eqref{effpotphotonorbitsgeneral}), which characterizes  the time-scale of the null-orbit instability, describes the damping of the black hole oscillations and therefore determines the imaginary part of the spectrum. We have 
\begin{equation}\label{eikonalspectrum}
\omega_{\text{QNM}}=\omegaR+i \omegaI= \Omega_m l-\text{i} \left(n+\frac{1}{2} \right)|\lambda|,
\end{equation} 
where 
\begin{subequations}
\begin{align}
&\Omega_m = \sqrt{\frac{A'(r)}{2r}}\biggl|_{r=r_m} = \frac{\sqrt{A(r_m)}}{r_m}; \label{Omegam}\\
&\lambda =\frac{1}{\sqrt{2}}\sqrt{-\frac{r_m^2}{A(r_m)}\left(\frac{d^2}{d\rst^2} \frac{A(r)}{r^2} \right)_{r=r_m}}. \label{LyapunovL}
\end{align}
\end{subequations}
Being $r_m$ the position of the peak of the geodesic potential, it is given by the solution of the \cref{minpot}. Although we have seen that, for extremal black holes, the potential $V$ has an additional minimum exactly at the horizon, we will not consider this contribution, as small perturbations from extremality have the effect of moving this minimum towards the black hole interior. The QNMs spectrum will be therefore entirely determined by the contribution at the maximum of $V$.

The general expressions \eqref{Omegam} and \eqref{LyapunovL} are valid for all spherically-symmetric, static and asymptotically-flat solutions, in the eikonal limit. For our general class of models, using \cref{f10,s1a,Omegam,LyapunovL}  we get
\begin{subequations}
\begin{align}
    &\omegaR=  \frac{l}{r_m}\sqrt{A(r_m)}=\frac{l}{\ell y_m}\sqrt{1-\alpha F_m},\label{s3}\\
    &\omegaI=  -\left(n+\frac{1}{2} \right)\frac{1}{\sqrt 2}\sqrt{A(r_m) r_m \biggl|\left( \frac{A'(r)}{r}\right)'_m\biggr| }=
     -\left(n+\frac{1}{2} \right)\frac{1}{\sqrt 2 \, \ell}\sqrt{\alpha \left(1-\alpha F_m\right)y_m \biggl|\left(\frac{F'}{y}\right)'_m\biggr|}\, .
    \label{s4}
\end{align}
\end{subequations}
It is important to stress that the QNMs frequencies depend both from the classical hair $M$ and from the quantum hair $\ell$. This dependence from two  parameters of the QNMs spectrum will have a well-defined signature in the  ringdown part of the  gravitational wave   generated in the   merging of  two compact objects  to form  a single black hole. Next-generation gravitational-wave detectors are expected to be sensitive enough to detect such effect.

In the generic case, $\omegaR$ and $\omegaI$ will be complicated functions of $\alpha$. Simpler expressions can be obtained for near-extremal black holes, by expanding in powers of $(\alpha- \alpha_c)$. Taking into account  that $|A|_m\neq 0$ (see the remark above), if we just  consider the near-extremal expansion but not the  near-horizon expansion, we will also have $\left(\frac{A(r)'}{r}\right)'_m\neq 0$. At first order in $(\alpha- \alpha_c)$ we get $\omegaI= \text{constant}/\ell + \text{constant} \ (G/\ell^2)(M-M_c)$ and similarly for $\omegaR$. Using \cref{qscalinggeneral}, we can express the quasi-normal frequencies in terms of the black hole temperature
\begin{subequations}
 \begin{align}
    \omegaR &= \frac{a}{\ell}+ b \, \ell \ T_\text{H}^2, \label{s5a}\\
    \omegaI &= \frac{c}{\ell}+ d \, \ell \ T_\text{H}^2 \label{s5},
\end{align}
\end{subequations}
where $a,b,c,d$  are  dimensionless constants. 

On the other hand, if we take   the near-horizon limit together with the near-extremal limit,  the metric satisfies $\left( \frac{A(r)'}{r}\right)'=0$, identically, since the geometry becomes that of AdS$_2\times$ S$^2$. While the behavior of $\omegaR$ remains that of \cref{s5a}, we get a linear scaling of $\omegaI$ with the temperature, owing to the absence of the constant term inside the square root:
\begin{equation}\label{ff1}
\omegaI\propto T_\text{H}.
\end{equation}

These results confirm only partially Hod's conjecture, which asserts the complete absence of the imaginary damped part in the spectrum in the near-extremal case, both for RN and Kerr black holes \cite{Hod:2008se,Hod:2008zz,Hod:2011zzd,Hod:2012zzb,Hod:2015hga}. In the case under consideration, Hod's conjecture seems to hold true only in the near-extremal, near-horizon case. This seems to be a feature of also general charged and rotating regular black-hole models \cite{Ansoldi:2006vg,Modesto:2010rv, Lan:2020fmn}. 

On the other hand, our results seem to confirm a general behavior found in Ref. \cite{Yang:2012pj} for the near-extremal Kerr space-time, which is characterized by a branching in the quasi-normal spectrum. One family, corresponding to the simple near-extremal limit, has indeed a non-vanishing imaginary part in the extremal limit, while the other branch shows that the damped part of the spectrum goes to zero in the near-extremal, near-horizon limit.

Finally, the temperature scaling \eqref{ff1} fully confirms previous derivation of the quasi-normal spectrum for two dimensional  AdS$_2$ black holes \cite{LopezOrtega:2011np, Cordero:2012je, Kettner:2004aw, Bhattacharjee:2020nul, Cadoni:2021qfn}. In fact, the latter allows for an explicit \textit{analytical} treatment through different methods, which all point towards the same result: a linear scaling of the imaginary part $\omegaI$ with the temperature of the hole. A quite interesting consequence of this scaling is the complete absence of the imaginary damped part in the spectrum in the extremal case, as the temperature becomes zero. These zero-damped (or nearly zero-damped) modes \cite{Hod:2008se,Hod:2008zz,Hod:2011zzd,Hod:2012zzb,Hod:2015hga,Zimmerman:2015trm,Joykutty:2021fgj} would therefore represent a clear phenomenological signature of extremal black holes. 

\section{ A minimal model: the Hayward  black hole} 
\label{sect:hayward}

The simplest example of our general class of models is given by the Hayward black-hole metric \cite{Hayward:2005gi,Frolov:2016pav}, for which the metric function $F$ in \cref{f10} reads 
\begin{equation}\label{haywbh}
F(y)= \frac{y^2}{y^3+1},
\end{equation}
with $y\ge 0$. As already mentioned in \cref{sect:general}, the analysis of Ref. \cite{Knorr:2022kqp} shows that an algebraic form of $F$ could be inconsistent with semiclassical field equations derived from an action principle,  at least if if one requires the solutions to be derived from an Einstein-Hilbert action with higher-order terms in the curvature. 

The horizon location and the extremality condition \eqref{f11} are now
\begin{equation}\label{haywbhext}
y^3-\alpha y^2+1=0,\quad -2\alpha y+ 3y^2=0.
\end{equation}
Solving these equations  yields the critical values of the parameter $\alpha$ and the critical radius $r_c$ 
\begin{equation}\label{haywbhextemal}
\alpha_c=\frac{3}{\sqrt[3]{4}},\quad r_c=\frac{2}{3}\RS=\sqrt[3]{2} \, \ell.
\end{equation}
The black hole has two horizons for $\RS> 3\ell/\sqrt[3]{4}$, is extremal for $\RS= 3 \ell/\sqrt[3]{4}$, whereas it becomes a horizonless star for $\RS< 3\ell/\sqrt[3]{4}$. The energy density $\epsilon$ and the mass function $m$  sourcing the   black  hole are  given by

\begin{equation}\label{haywbhMepsilon}
\epsilon(r)=\frac{3}{4\pi}\frac{M \ell^3}{(r^3+\ell^3)^2},\quad   m(r)=\frac{M r^3}{r^3+\ell^3}.
\end{equation}
The NEC  \eqref{NullEnergyCondGeneral} is always satisfied, while the SEC \eqref{StrongEnergyCondGeneral} is violated deep inside the core of the object, for $r \leq \ell/\sqrt[3]{2}$. On the other hand, for $\ell\to 0$, or for $r\to\infty$, $\epsilon(r)$  has a very small, Dirac's delta-like support only in the region near $r=0$, and therefore $m(r)=M$: the Schwarzschild black hole is recovered. The deviation from the Schwarzschild case can be characterized by defining a mass deviation $\Delta m(r)$ as the difference between the mass at infinity $M$ and $m(r)$, which in the present case reads
\begin{equation}\label{haywardDeltam}
\Delta m \equiv M-m(r) = \frac{M \ell^3}{r^3+\ell^3}.
\end{equation}
For $r \to \infty$, it behaves  as
\begin{equation}\label{HaywardAsymptDeltam}
\Delta m \sim \frac{M\ell^3}{r^3}+\mathcal{O}\left(\frac{1}{r^4} \right).
\end{equation}

The solution is asymptotically flat and satisfies the boundary conditions $\epsilon$, $p_\parallel$, $p_\perp \to 0$ for $r \to \infty$, while it has a dS behavior near $r=0$ with the dS length  $\hat L$ \eqref{massconstraint}  given by
\begin{equation}\label{haywbhL}
\hat L=  \ell^{3/2} \RS^{-1/2},
\end{equation}
which relates $\ell$ with the Schwarzschild radius and the dS length characterizing the small $r$ behavior.
\Cref{haywbhL} fully confirms  the validity of our general scaling given by \cref{f4}. 

We note that the same result in \cref{haywbhL} can be obtained in the limit of very large $\ell$. In this case, however, we have an \textit{exact} solution of Einstein's equations, sourced by a constant-density, \textit{isotropic} and \textit{homogeneous} fluid, with equation of state
\begin{equation}\label{dSlimitr0}
p_\parallel=p_\perp = -\epsilon = -\frac{3}{8\pi G \hat{L}^2}.
\end{equation}
Indeed, looking at the density profile \eqref{haywbh}, the dS universe can be recovered in the limit $\ell \to \infty$ only if $M \to \infty$, so that the energy density of the source \eqref{haywbh} becomes  constant. \\

For $\alpha>\alpha_c$, the cubic equation in (\ref{haywbhextemal}) has three real roots, out of which two are positive, $r_+$ and $r_-$, denoting the outer and inner horizons respectively, whereas the third, $r_3$, is negative. The metric function $A$ factorizes as
\begin{equation}\label{haywbhgeneral}
A(r)=  \frac  {(r-r_+)(r-r_-)(r-r_3)}{r^3 + \ell^3}.
\end{equation}
In the extremal limit $r_+=r_-\equiv r_c=(2/3)\RS$, after a translation of the radial coordinate $r\to r +r_c$, the space-time metric becomes    
\begin{equation}\label{me1}
ds^2=  -\tilde A(r)dt^2+ \tilde A^{-1}(r)  dr^2 + (r+r_c)^2 d\Omega^2,\quad \tilde A(r) \equiv  \frac{r^2(r+r_c-r_3)}{(r+r_c)^3+\ell^3},
\end{equation}
describing an asymptotically flat region connected with an infinitely long throat of radius $r_c$. The near horizon (around $r=0$) expansion of the metric \eqref{me1} gives the AdS$_2\times$ S$^2$ space-time \eqref{f51}, with an AdS$_2$ length $L_2= (2/3)\RS$. The fluid stabilizing the extremal solution is characterized by the equations of state \eqref{epsilonpressuresnearhorgeneral}, where now  the AdS$_2$   length $L_2$  and the radius $r_c$ of the two-sphere have  the same  value, $ L_2=r_c= (2/3)\RS$.   
\subsection{Thermodynamics  and phase transition}
Inserting $F$ given by \cref{haywbh}  into \cref{THQBHgeneral}, we get the mass and temperature of the black hole
\begin{equation}\label{tm2}
T_\text{H}=  \frac{1}{4\pi\, \rH}\frac{\rH^3-2\ell^3}{\rH^3+\ell^3},\quad M=  \frac{1}{2G}\left(\rH+ \frac{\ell^3}{\rH^2}\right).
\end{equation}
The temperature behavior agrees with the expected qualitative one shown in \Cref{Temp1}. The maximum of $T_\text{H}$ is obtained by solving the equation $y^6-10y^3-2=0$, giving $\rH=\sqrt[3]{(5+3\sqrt 3)} \,\ell$.  

Expanding \cref{tm2} near extremality, one easily finds the quadratic scaling  \eqref{qscalinggeneral} of the mass with the  temperature   
\begin{equation}\label{qscaling1}
 M-M_c=12 \pi^2 \frac{\ell^3}{G}T_\text{H}^2.
\end{equation}

The black-hole entropy satisfying the first principle of thermodynamics is easily obtained from \cref{gg3}
\begin{equation}\label{tm3}
S=  \frac{\pi  \rH^2}{G}-\frac{2\pi\ell^3}{G \rH}
\end{equation}
The first term is the standard area law, while the second term describes  $\ell$-dependent deviations.

\begin{figure}
\centering
\includegraphics[width= 8 cm, height = 8 cm,keepaspectratio]{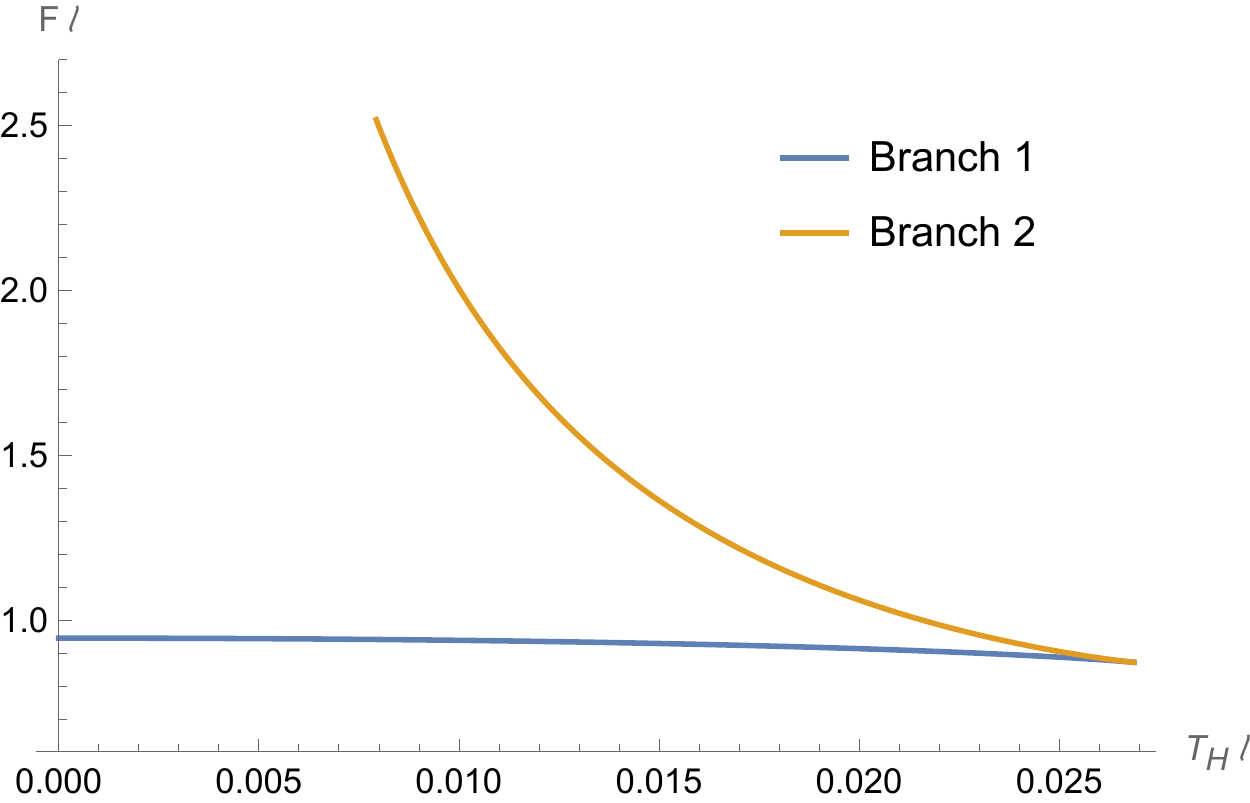}
\caption{Free energy $\mathcal{F}$, in units of $\ell^{-1}$, as a function of the temperature, in units of $\ell^{-1}$,  for the  two branches of  the Hayward black hole. We plot    $\mathcal{F}$  for branch $I$,  $\ell \sim \RS$ (solid blue curve), and  for branch $II$  (solid orange curve) corresponding to $\ell \ll \RS$.  We see that  "quantum deformed" black holes with $\ell \sim\RS$  are always energetically preferred  with respect to  those with  $\ell \ll \RS$.} 
	\label{PhaseDiagramHayward}
\end{figure}

The specific heat $C$ can easily be calculated using \cref{tm2} and agrees with the qualitative behavior shown in \Cref{SpeHeat1}. It diverges for $\rH=\sqrt[3]{(5+3\sqrt 3)}\, \ell$, indicating the onset of the second order phase transition, with the stable thermodynamic branch I occurring for $\rH<\sqrt[3]{(5+3\sqrt 3)}\, \ell$ \footnote{The possibility of having a phase transition in the Hayward model has been previously recognized in Ref. \cite{Molina:2021hgx}}.

The existence of this phase transition and  related thermodynamical phase portrait  can  be checked  by computing    the  
free energy $\mathcal{F}= M-T_\text{H} S$  as a function of the temperature.  The free energy  for the two branches $I$ and $II$ has to be calculated numerically by inverting the equation $T_\text{H}=T_\text{H}(\rH)$. We plot  $\mathcal{F}(T_\text{H})$ in \Cref{PhaseDiagramHayward}. Branch $I$ corresponds to $r_\text{H}$ varying between the extremal value $r_c$ and $r_{\text{H, max}}$, and thus it describes quantum black holes for which the quantum deformation parameter $\ell$ is of the same order of magnitude of the classical Schwarzschild radius $\RS$. Conversely, branch $II$ corresponds to $r_\text{H}$ taking values much larger than $r_{\text{H, max}}$, where $\ell \ll \RS$, corresponding to the classical black-hole branch.\\
We note that the branch $I$ is always energetically preferred with respect to branch II, which further supports our  choice $\ell \sim \RS$.

\subsection{Null geodesics and quasi-normal modes in the eikonal limit}
Let us now consider   geodesic motion and  QNMs for the Hayward  black hole.
The effective potential \eqref{effpotphotonorbitsgeneral}, which determines photon orbits in the black-hole background, in the present case reads
\begin{equation}\label{effpotphotonorbits}
V(r) = \frac{J^2}{2 r^2}\left(1-  \frac{\RS r^2}{r^3+\ell^3}\right).
\end{equation}
The extrema of $V(r)$ are determined by \cref{minpot} with $A$ given by \cref{haywbhgeneral}, i.e. by the roots of the equation $2y^6-3\alpha y^5+4 y^3+2=0$. By solving this equation numerically, one can show that no real roots exist for $\alpha<\alpha_m\approx 1.64$. The position of the maximum of the potential will depend on the value of the parameter $\alpha$ with respect to  $\alpha_m$ and  $\alpha_c=3/\sqrt[3]{4}$. 
Also here, we can distinguish between the four cases shown in \Cref{photonOrbgeneral}.
The quasi-normal frequencies in the eikonal approximation for the  Hayward black-hole   can be easily calculated plugging \cref{haywbh} into \cref {s3,s4}. One has 
\begin{equation}\label{s2a}
\omegaR=  \frac{l}{\ell y_m} \sqrt{1-\alpha \frac{y^2_m}{y^3_m+1}},\quad  \omegaI=-\left(n+\frac{1}{2} \right)\frac{\sqrt{3\alpha} \, y_m^{3/2}}{\sqrt{2} \, \ell \left(1+y_m \right)^2}\sqrt{|y_m^3-5| \left(1+y_m^3-\alpha y_m^2 \right)}.
\end{equation}
By expanding around $\alpha = \alpha_c = 3/\sqrt[3]{4}$, we get
\begin{subequations}
\begin{align}
&\omegaR \simeq \frac{0.21\, l}{\ell}-\frac{0.11 \ l}{\ell}\left(\alpha-\alpha_c \right)\simeq \frac{0.21 \, l}{\ell}-26.13 \, l \ \ell \ T^2_\text{H}; \\
&\omegaI \simeq -\left(n+\frac{1}{2} \right) \frac{5.70}{\ell}+\left(n+\frac{1}{2} \right)\frac{1.44}{\ell}\left(\alpha-\alpha_c \right) \simeq -\left(n+\frac{1}{2} \right) \frac{5.71}{\ell} + \left(n+\frac{1}{2} \right) 340.07\ \ell \ T^2_\text{H},
\end{align}
\end{subequations}
where in the last equalities we used the definition of $\alpha$ and \cref{qscaling1}.

\section{Quantum black holes with gaussian cores}
\label{sect:gaussian}

Another explicit and simple realization of a regular quantum black hole, sourced by an anisotropic fluid with EoS given by \cref{EoSdarkenergy} and satisfying the conditions outlined in \Cref{sect:general} , can be obtained by taking a gaussian density distribution in the interior of the astrophysical object,  peaked at $r=0$
\begin{equation}\label{gaussianprofile}
\epsilon(r) = \frac{M}{\pi^{3/2}\ell^3} e^{-r^2/\ell^2},
\end{equation}
where $M$ is the MS mass as seen from infinity, namely
\begin{equation}
M = 4 \pi \int_0^{\infty} dr \ r^2 \epsilon(r).
\end{equation}
The parameter $\ell$ represents  here  a smearing of the classical Schwarzschild-Dirac delta-density profile (the latter is recovered in the limit $\ell \to 0$). Such a density profile is motivated by several microscopic descriptions of black holes: non-commutative geometry \cite{Nicolini:2005vd, Ansoldi:2008jw}, loop quantum gravity (LQG) \cite{Modesto:2010uh}, corpuscular picture \cite{Casadio:2013hja}. The NEC condition \eqref{NullEnergyCondGeneral} is always satisfied, while the SEC \eqref{StrongEnergyCondGeneral} is again violated in the deep core of the body, i.e. for $r < \ell$. \\

Near $r =0$, the space-time behaves as dS,  with a dS length $\hat L$  given by
\begin{equation}
\label{k1}
\hat L= \sqrt{\frac{3}{4}}\pi^{1/4} \, \ell^{3/2} \RS^{-1/2},
\end{equation}
confirming again our general result given by \cref{f4}.
 
As we also saw in \cref{sect:hayward}, the same result can be obtained as an exact solution of Einstein's field equations, sourced by the fluid with EoS \eqref{dSlimitr0}, in the limit $\ell \to \infty$. Indeed, looking at the density profile \eqref{gaussianprofile}, the dS universe can  be recovered in the limit $\ell \to \infty$ only if $M \to \infty$, and the energy density \eqref{gaussianprofile} behaves as a constant. 

Let's turn our attention to the metric structure, by computing the MS mass at a generic $r$
\begin{equation}\label{MSmassgaussian}
\begin{split}
m(r) &= 4 \pi \int_0^r d\tilde r \ \tilde r^2 \ \epsilon(\tilde r)=  \frac{2M}{\sqrt{\pi}} \gamma\left(\frac{3}{2},  \frac{r^2}{\ell^2} \right)= M\left[1 -\frac{2}{\sqrt{\pi}} \Gamma\left(\frac{3}{2},  \frac{r^2}{\ell^2} \right)\right]
\end{split}\end{equation} 
where $\gamma(a, z) =  \int_0^z dt \ e^{-t} \, t^{a-1}$ and $\Gamma(a,z)= \int_z^\infty dt \ e^{-t} \, t^{a-1}$ are the incomplete gamma functions. The first term in \cref{MSmassgaussian} is the mass measured at infinity (the Schwarzschild ADM mass), while the second term, parametrized by $\ell$, encodes the quantum corrections, the effects of the smearing of the singularity. The deviation from the Schwarzschild solution, described by the mass deviation $\Delta m(r) \equiv M-m(r)$, is strongly suppressed outside the core of the compact object, as for $r \to \infty$ it behaves  as
\begin{equation}\label{DeltaMGaussian}
\Delta m \sim  \frac{M}{\sqrt{\pi}}e^{-\frac{r^2}{\ell^2}} \left(\frac{\ell}{r}+\frac{2  r}{\ell}  \right)
\end{equation}
which represents a stronger suppression with respect to that of the Hayward model (see \cref{HaywardAsymptDeltam}).

The metric components can be written in the form of \cref{f10} with the metric function  $F(y)$ given by

\begin{equation}\label{metricgaussian}
F(y)=\frac{1}{y}\left[1 -\frac{2}{\sqrt{\pi}} \Gamma\left(\frac{3}{2}, y^2 \right) \right].
\end{equation}

 Contrary to the Hayward model, in the present case, the non-algebraic form of $F$ allows to circumvent the viability constraints of Ref. \cite{Knorr:2022kqp}.

Given the  form of the metric functions, the position of the horizon(s) and the parameters range discriminating between the two-, one- or no horizons  cases  have to be computed numerically. The two horizons merge and we have the extremal case when both equations in \eqref{f11} are satisfied. This translates into the conditions:
\begin{subequations}
\begin{align}
& 1-\frac{2GM }{y \ \ell}+ \frac{4 G M}{\ell \sqrt{\pi} \ y} \Gamma\left(\frac{3}{2}, y^2 \right)=0; \label{g00y} \\
&1-\frac{2}{\sqrt{\pi}} \Gamma\left(\frac{3}{2},y^2 \right)-\frac{4 y^3}{\sqrt{\pi}}e^{-y^2}=0. \label{g00primey}
\end{align}
\end{subequations}
These equations need to be solved numerically \footnote{We discard the simplest analytical solution, i.e. $r=0$, since it is not a minimum.}. The numerical solution of \cref{g00primey} is $y_\text{min} \simeq 1.51$, which means $r_\text{min} \simeq 1.5 \, \ell $. The range of parameter $\alpha=\RS\ell^{-1}$ discriminating between the aforementioned three cases therefore is

\begin{itemize}
\item {\bf No horizons}: $A\left(y=y_\text{min}\right) > 0$
\begin{equation}\begin{split}
\alpha < \frac{3}{2-\frac{4}{ \sqrt{\pi}}\Gamma \left(\frac{3}{2},\frac{9}{4} \right)}.
\label{boundnohor}
\end{split}\end{equation}
\item {\bf One horizon}: $A\left(y=y_\text{min}\right) = 0$
\begin{equation}\label{Monehorizon}
\alpha  = \frac{3}{2  - \frac{4}{\sqrt{\pi}}\Gamma\left(\frac{3}{2},\frac{9}{4} \right)}\simeq 1.90.
\end{equation}
\item {\bf Two horizons}: $A\left(y=y_\text{min}\right) > 0$
\begin{equation}\label{hl1}
\alpha  >\frac{3}{2  - \frac{4}{\sqrt{\pi}}\Gamma\left(\frac{3}{2},\frac{9}{4} \right)}.
\end{equation}
In this latter case, the outer horizon is an event horizon, while the inner one is a Cauchy horizon \cite{Ansoldi:2008jw}.
\end{itemize}

We have therefore a critical value $\ell_c$ for the quantum parameter $\ell$ for which the two horizons merge and the quantum black hole becomes extremal 
\begin{equation}\label{g1}
 \ell_{c} \simeq 0.53 \RS,
\end{equation}
which is  close to the classical gravitational radius of the compact object. This critical value discriminates between the three classes of solutions. When $\ell$ is above, equal or below $\ell_c$ we have a solution with two, one or none horizons, correspondingly. $\ell_c$, in turns, determines a critical value $\hat L_c$ for dS length through \cref{k1}, which also turns out to be close to the classical Schwarzschild radius \footnote{The fact that $\hat L_c < \ell_c$ is expected: the SEC is violated in the deep core of the object, namely for $r<\ell$. This is perfectly consistent with the fact that this energy condition is violated in a dS space-time.}: 
  \begin{equation}\label{g3}
 \hat L_c \simeq  0.31 \RS.
\end{equation}

The  most interesting case is the  extremal black hole, obtained for   $\ell=\ell_c$ ($\hat L=\hat L_c$).  As discussed in sect. \ref{sect:general}, in the extremal  case the near-horizon geometry factorizes  as AdS$_2\times$ S$^2$, with the AdS$_2$ length  given by  $L_2^{-1}=\sqrt{-\frac{1}{2}A(r_c)''}$. 
$L_2$  can be calculated first using \cref{g00y,g00primey} to get
\begin{equation}\label{s1}
e^{-y^2(r_c)}=\frac{\sqrt{\pi}  \ell}{4\RS y^2(r_c)}.
\end{equation}
Using this equation together with \cref{g00y,g00primey,MSmassgaussian,metricgaussian} enables us to find 
\begin{equation}\label{s2}
-\frac{1}{2} A(r_c)''\equiv \frac{1}{L_2^2}= \frac{1}{\ell^2}- \frac{1}{r_c^2} .
\end{equation}
Using \cref{g1} and taking into account that $r_\text{min} \simeq 1.5 \, \ell\simeq 0.78 \, \RS$ we get
\begin{equation}\label{h3}
L_2\simeq 0.70 \, \RS,
\end{equation}
confirming  the general result of Sect.  \ref{sect:general}, according to which   both  the radius of the two-sphere and the AdS$_2$ length are of  order $\RS$. 

As shown at the end of \cref{sec:ExtremalLimitGeneral}, the extremal solution is stabilized by a particular profile for the energy density $\epsilon$ and for the pressures $p_\parallel$   and   $p_\perp$, given by the expressions \eqref{epsilonpressuresnearhorgeneral}. In the present case, the negative energy density sourcing the AdS$_2$ space-time reads $\epsilon^{\text{(nh)}}_{\text{(AdS)}}\equiv  -1/(8\pi G L_2^2) =-(1/8\pi G)\left(1/\ell^2- 1/r_c^2\right)$.

\subsection{Quantum black hole regimes}
\label{sec:QuantumBHregimes}

The parameter $\ell$ (or equivalently  $\hat L$) controls the scale of quantum effects in our quantum black-hole model. In the usual, most conservative approach, which assumes quantum gravity effects to be relevant only at the Planck scale $\lP$, $\ell$ is assumed to be of the same order of magnitude of $\lP$. This assumption is surely justified when $\ell$ has an explicit origin in the microscopic description of gravity at the Planck scale. It is for instance the case of Refs. \cite{Nicolini:2005vd, Ansoldi:2008jw, Casadio:2013hja,Modesto:2010uh}, where the gaussian model, and the parameter $\ell$ in particular, parametrizes UV non-commutative \cite{Nicolini:2005vd} or LQG \cite{Modesto:2010uh} effects.

However, this is not the case in those approaches, like the one followed in this paper, in which an IR quantum scale, hierarchically smaller than $\lP$, is generated. Given the attention the model with $\ell \sim \lP$ received in the past \cite{Modesto:2010uh, Nicolini:2005vd, Spallucci:2011rn,Modesto:2004xx, Modesto:2008im, Nicolini:2008aj, Hossenfelder:2009fc,Sprenger:2012uc,Bambi:2013caa,Casadio:2014pia}, it is worthwhile to quantitatively compare the two different regimes $1)$  $\ell\sim \lP$ and $2)$ $\ell\sim \RS$ for the black-hole model with a gaussian core, and analyze the possible impact on observable phenomenology in the two cases. Again, we remind that we are considering macroscopic black holes, i.e $\RS\gg  \lP $.

\subsubsection*{$\ell\sim \lP$}
In this case, \cref{hl1} tells us that we always have two horizons, the black  hole  is far from extremality and the inner horizon is very close to $r=0$. It is quite easy to understand that these quantum effects are completely irrelevant for macroscopic black holes, at least for what concerns  the  phenomenology  accessible  to external observers. In fact, the matter density is sensibly different from zero only at distances of order $\lP$ from the classical singularity at $r=0$. Therefore, for the external observer, the deviations from the Schwarzschild solution are expected to be incredibly small.  We have sensible deviations from $\rH \sim \RS$ only for black holes with masses of order the Planck mass $m_\text{P} = 1/\sqrt{G}$, where the event horizon is slightly less than $\RS$ (the radius of the outer horizon is $\rH \simeq 1.8 \sqrt{G} =0.9 \RS$). However, as the mass increases, the outer horizon becomes rapidly indistinguishable from $\RS$. For example, for a solar mass black hole, $M = 1 \ M_\odot$, the outer horizon of the metric \eqref{metricgaussian} is at $\RS$ and the corrections are exponentially suppressed by a factor $\Gamma\left(\frac{3}{2},\frac{\rH^2}{\lP^2} \right)\sim e^{-\rH^2/\lP^2}=e^{-\RS^2/\lP^2}= e^{-\frac{4 G M^2_\odot}{c \hbar}}\sim e^{-10^{76}}$  (note that, in the last expression, we have reinstated the speed of light $c$ and the Planck constant $\hbar$). The mass deviation at the horizon $\Delta m(\rH)$ is of the same order of magnitude, which is effectively zero from the point of view of the external observer.

\subsubsection*{$\ell\sim \RS$}

As explained in the introduction, there are several indications pointing at the relevance of quantum effects at horizon scales. However, presently we do not have a precise microscopic description of these quantum effects, but only some quite interesting proposals, like e.g. fuzzballs, firewalls, non-local effects and corpuscular models. An interesting explicit corpuscular black-hole model with a gaussian core is the one given in Ref. \cite{Casadio:2013hja}, whose density profile is 
\begin{equation}\label{densityprofileCO}
\epsilon_{\text{corpuscular}} = \frac{7^2 m_\text{P}}{\sqrt{\pi}N}\frac{e^{-\frac{7}{2}\frac{r^2}{N \lP^2}}}{\lP^3}
\end{equation}
where $N$ is the number of gravitons building  up the black hole. Comparing this profile with our model \eqref{gaussianprofile}, we can read the values of our  parameters $\ell$, $M$ in terms of $N$ 
\begin{equation}
\ell = \lP\sqrt{\frac{2N}{7}}, \qquad M = 49 \pi\frac{\ell^3}{N \lP^4}.
\end{equation}
From these equations, one easily gets  the expected  holographic scaling  of  $N$,   $N  \propto \lP^2 M^2$ and a value of $\ell$ which is $\ell= \RS/28 \pi \simeq   0.01 \RS<\ell_c$. The black hole has two horizons and is far from extremality. The  outer  horizon is quite close to the classical Schwarzschild one, we have $\rH\simeq  0.96\RS$. Therefore, the mass deviation is again quite small:
\begin{equation}
\frac{\Delta m}{M}\biggr|_\text{H} = \frac{2}{\sqrt{\pi}}\Gamma\left(\frac{3}{2},y_\text{H}^2 \right)\sim e^{-7733}.
\end{equation}
We see that, for a value of $\ell$ which is about $1/100$ of the critical value $\ell_c$, deviations from the classical behavior are still quite small. \\

As a last example we consider a  value   $\ell<\ell_c$, but quite close to the critical value, $\ell = \ell_c/2$. The space-time  has  two horizons  and the outer one is at
\begin{equation}
r_\text{H}\simeq 0.92 \RS,
\end{equation}
which is a small, but still important, difference with respect to the classical radius $\RS$. The mass deviation is 
\begin{equation}
\frac{\Delta m}{M}\biggr|_\text{H} = \frac{2}{\sqrt{\pi}}\Gamma\left(\frac{3}{2}, \frac{c^4 r^2_\text{H}}{G^2 M^2} \right)\simeq  \frac{2}{\sqrt{\pi}}\Gamma\left(\frac{3}{2}, 1.84^2 \right)\simeq 0.07.
\end{equation}
These results further show that the most interesting regime is that for which $\ell\sim \RS$ from both a purely theoretical and from a phenomenological points of view.

\subsection{Thermodynamics and phase transition}

\begin{figure}
\centering
\includegraphics[width= 8 cm, height = 8 cm,keepaspectratio]{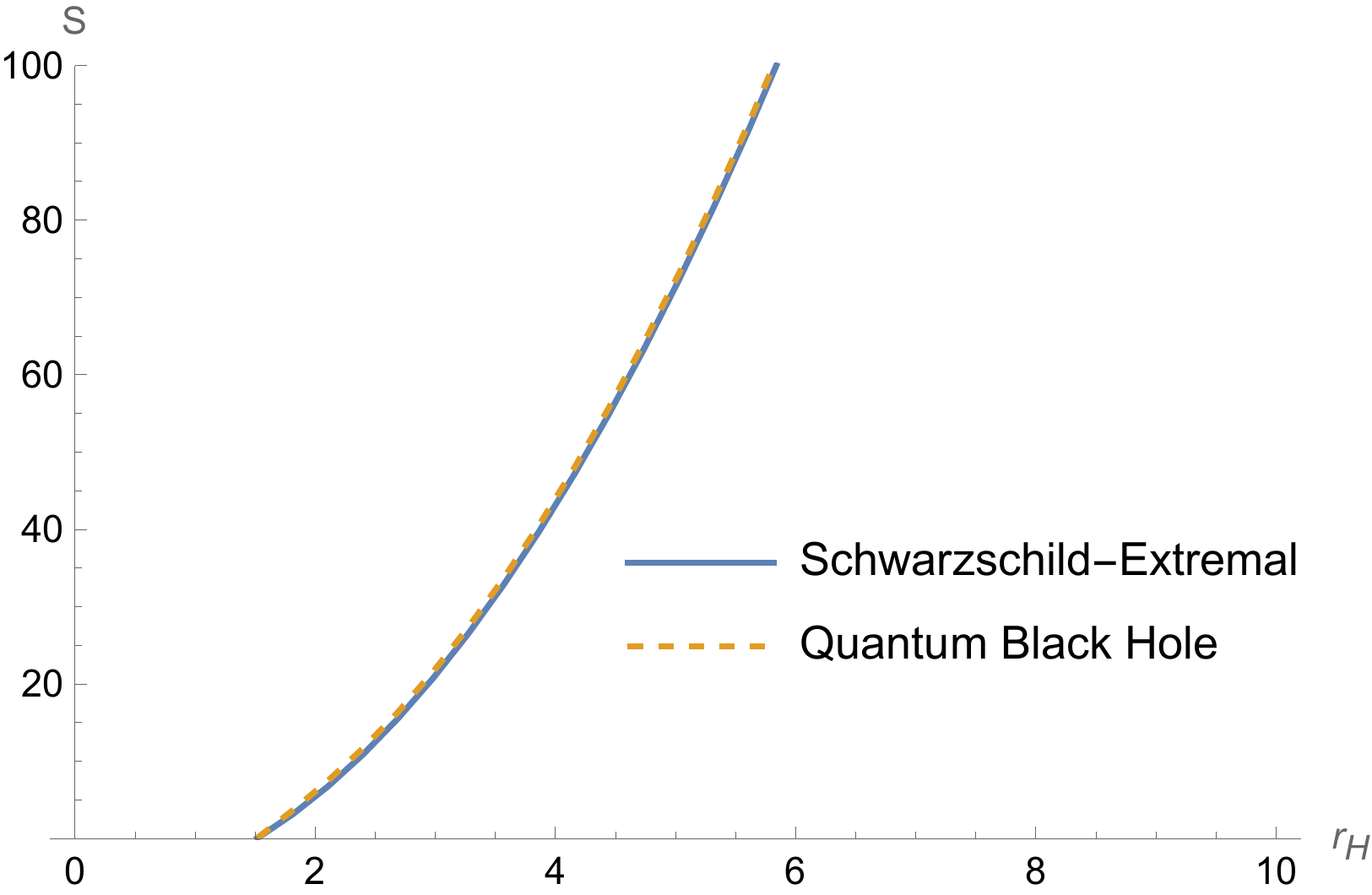}
\caption{Plot  of the entropy of the  quantum gaussian black hole (dashed orange line) vs standard Hawking entropy of the Schwarzschild black hole (minus the corresponding Hawking entropy of the extremal case). For simplicity, we set $\ell =1$ and $G=1$.}
	\label{Entropy}
\end{figure}

Inserting the metric function \eqref{metricgaussian} into eqs. \eqref{THQBHgeneral}, we get the temperature and the ADM mass of the gaussian black holes 
\begin{equation}\label{THQBHexplicit}
T_\text{H}=\frac{1}{4\pi r_\text{H}} \left[1- \frac{8 GM(r_\text{H})r^2_\text{H}}{\ell^3}e^{-r^2_\text{H}/\ell^2} \right], \quad M(\rH)= \frac{\rH}{2G\left[1-\frac{2}{\sqrt{\pi}}\Gamma\left(\frac{3}{2},\frac{\rH^2}{\ell^2} \right) \right]}.
\end{equation}
The temperature is  given by the standard Hawking result plus an $\ell$-dependent term, encoding quantum deviations from standard black hole thermodynamics, which however are exponentially suppressed. The behavior agrees with the qualitative one depicted in \Cref{Temp1}. It starts from zero in correspondence with the extremal case. Then, it rises and reaches a maximum, whose position is given by solving the equation $dT_\text{H}/d\rH=0$, which in the present case is at $r_{\text{H, max}}\simeq 2.38 \ell$. Finally, it decreases and reaches zero as $\rH/\ell \to \infty$, in agreement with the fact that we have to recover the standard Schwarzschild results $T_\text{H}=\frac{1}{4\pi r_\text{H}},\,\, M=\frac{r_\text{H}}{2G}$ in this limit. Also in this case, the quantum deformation  parameter  $\ell\neq 0$ solves the singular thermodynamic behavior of the Schwarzschild temperature $T_{\text{H}}\to \infty$ for $r_{\text{H}}\to 0$. 
 
On the other hand, from \cref{g00y}, we get the value of the deformation parameter corresponding to $r_{\text{H, max}}$ 
 \begin{equation}\label{df1}
\ell_{\text{H, max}}=\frac{1-\frac{2}{\sqrt{\pi}}\Gamma\left(\frac{3}{2},y^2_\text{H, max} \right)}{y_\text{H, max}} \RS \simeq 0.42 \RS.
\end{equation}

The entropy of the black hole can be obtained by integrating \cref{Entropydiffgeneral} numerically, using the fact that  the zero-entropy state is at the extremal radius $r_\text{H, extremal}/\ell \simeq 1.51$. The result of the integral is presented in \Cref{Entropy} (dashed orange line) and is compared to the standard result for the Schwarzschild black hole \footnote{In this case, the blue solid curve in \Cref{Entropy} is obtained by subtracting the Hawking entropy of the extremal configuration $S_\text{extremal}=\pi r^2_\text{H,extremal}$ (with $G=1$) from the standard Schwarzschild entropy $\pi\rH^2$.}. As it can be seen, the entropy does not differ  significantly from the standard area law, as quantum deviations are expected to be exponentially suppressed \cite{Myung:2006mz,Banerjee:2008du,Nozari:2008rc} (see also \cref{DeltaMGaussian}). 

\begin{figure}
\centering
\includegraphics[width= 8 cm, height = 8 cm,keepaspectratio]{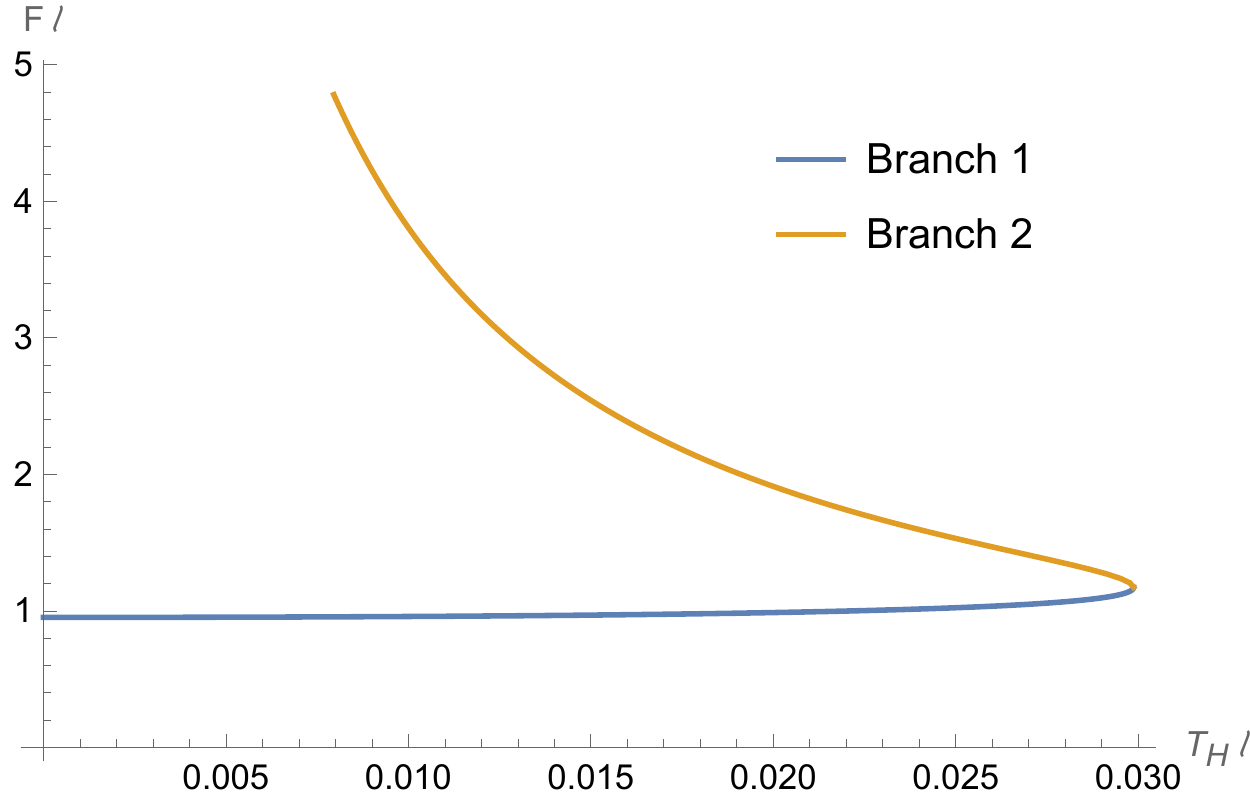}
\caption{Free energy $\mathcal{F}$, in units of $\ell^{-1}$, as a function of the temperature in units of $\ell^{-1}$, for the gaussian model. We distinguish between two branches: one corresponding to black holes with $\ell \sim \RS$ (solid blue curve), the other (orange curve) corresponding to $\ell \ll \RS$.}
	\label{PhaseDiagram}
\end{figure}

Expanding \cref{THQBHexplicit} near extremality yields the quadratic scaling \eqref{qscalinggeneral} of the mass above extremality with the temperature
\begin{equation}
M-M_c\simeq 15.55 \frac{\ell^3}{G}T_\text{H}^2.
\end{equation}
The specific heat can be computed using \cref{sh} and follows the qualitative behavior of \Cref{SpeHeat1}: it diverges at $r_{\text{H, max}}\simeq 2.38 \, \ell$, indicating the onset of the second-order phase transition. Indeed, by computing numerically the free energy $\mathcal{F}=M-T_\text{H} S$ and expressing it as a function of $T_\text{H}$, we get the phase diagram depicted in \Cref{PhaseDiagram}.

Again, we have two  branches. Branch $I$ corresponds to $r_\text{H}$ taking values  between the extremal value  $r_c$ and   $r_{\text{H, max}}$ (correspondingly, the parameter $\ell$ takes values between $ \ell_{\text{H, max}}\simeq  0.42 \RS$ and $\ell_c  \simeq 0.53\RS$), and therefore describes quantum black holes with $\ell \sim \RS$. Conversely, branch $II$ corresponds to $\rH$ much larger than $r_{\text{H, max}}$, corresponding to small ($<0.42 \RS$) values for the parameter $\ell$. Thus, the far right region of branch $II$ describes again classical black holes ($\ell\ll \RS$).

\Cref{PhaseDiagram} shows  that the free energy of branch $I$ is always smaller than that of branch $II$. This means that black holes in branch $I$, i.e. black holes with $\ell\sim \RS$, are always energetically preferred.   

\subsection{Null geodesics and photon orbits}
\label{sec:PhotonOrbitsGaussian}

The effective potential \eqref{effpotphotonorbitsgeneral} determining photon orbits in the gaussian quantum black hole reads

\begin{equation}\label{effpotphotonorbitsgaussian}
V(r) =\frac{J^2}{2 r^2} \left[1-\frac{2GM}{r}+\frac{4 GM}{\sqrt{\pi}r} \Gamma\left(\frac{3}{2}, \frac{r^2}{\ell^2} \right) \right],
\end{equation}
and its behavior for different values of  the parameter $\alpha \equiv \RS\ell^{-1}$ agrees with the qualitative plots shown in  \Cref{photonOrbgeneral}. 

The extrema of the potential are given by the zeros of $dV(r)/dr$  (see  \cref{minpot}), i.e. by the solution of the equation
\begin{equation}\begin{split}\label{photonradiuseq}
&-r + 3 GM -\frac{4 GM}{\sqrt{\pi}}\frac{r^3}{\ell^3} e^{-r^2/\ell^2}-\frac{6GM}{\sqrt{\pi}}\Gamma\left(\frac{3}{2},\frac{r^2}{\ell^2} \right)=0.
\end{split}\end{equation}
We see that the presence of $\ell$ shifts the position of the photon sphere from the Schwarzschild case $r=3GM$.\\
By numerically solving the equation above, we distinguish again between the three cases (2, 1 or no horizons)
\begin{itemize}
\item {\bf Two horizons} ($\alpha>\frac{3}{2  - \frac{4}{\sqrt{\pi}}\Gamma\left(\frac{3}{2},\frac{9}{4} \right)}$): we have multiple zeros, corresponding to one minimum in the black-hole interior and one maximum in the exterior, which corresponds to the position of the unstable photon orbit. We are interested in the latter. For instance, for $\alpha=2$ ($\ell = G M$ therefore), we have $y_m\simeq 2.99$, which corresponds to a position of the photon ring $r_m\simeq 2.99 \ell = 2.99 GM$, very close to the Schwarzschild case $3 GM$.
\item {\bf One horizon} ($\alpha=\frac{3}{2  - \frac{4}{\sqrt{\pi}}\Gamma\left(\frac{3}{2},\frac{9}{4} \right)}$): this case is similar to the previous one, we have a minimum (coinciding with the event horizon), which represents a stable photon orbit, and a maximum in the potential, representing the unstable photon orbit. Focusing again on the latter, we have $y_m\simeq 2.84$, corresponding to $r_m\simeq 2.98 GM$, again pretty close to the Schwarzschild result. 
\item {\bf No horizons} ($\alpha<\frac{3}{2  - \frac{4}{\sqrt{\pi}}\Gamma\left(\frac{3}{2},\frac{9}{4} \right)}$): we have two possible cases. If $\alpha$ is large enough ($\alpha \sim [1.6, 1.8]$), we can have a maximum corresponding to a photon ring. For example, for $\alpha =1.6$, we get $y_m\simeq 2.19$, which means $r_m\simeq 2.74 G M$, which is a significant departure from the standard result. However, if $\alpha$ falls below the aforementioned interval, we do not have a maximum anymore and the  photon ring cannot  be present. 
\end{itemize}

\subsection{Quasi-normal modes spectrum in the eikonal limit}
We can  use the general arguments and results of \Cref{sect:qnm} to compute the expression of the quasi-normal frequencies 
for the quantum   gaussian black hole in the eikonal limit. Applying the general formulae \eqref{s3}, \eqref{s4} to the case of \cref{metricgaussian}, we get
\begin{subequations}
\begin{align}
&\omegaR=\frac{l}{\ell y_m}\sqrt{1-\frac{\alpha}{y_m}\left[1-\frac{2}{\sqrt{\pi}}\Gamma\left(\frac{3}{2}, y_m^2 \right) \right]}\\
&\omegaI=-\left(n+\frac{1}{2} \right)\frac{1}{\sqrt{2}\ell} \sqrt{\alpha \left[y_m-\alpha + \frac{2}{\sqrt{\pi}}\Gamma\left(\frac{3}{2}, y_m^2 \right) \right] \biggl|\frac{3}{y_m^4}-\frac{4}{\sqrt{\pi}}\frac{1+2y_m^2}{y_m}e^{-y^2_m}-\frac{6}{\sqrt{\pi}y^2_m}\Gamma\left(\frac{3}{2}, y_m^2 \right) \biggr|}
\end{align}
\end{subequations}
By expanding around the critical value $\alpha_c$ \eqref{Monehorizon}, we get
\begin{subequations}
\begin{align}
&\omegaR\simeq \frac{l}{\ell}\left[0.20-0.10(\alpha-\alpha_c) \right]\simeq 0.20 \frac{l}{\ell}-3.36 \, l \ \ell \ T_\text{H}^2; \\
&\omegaI\simeq \frac{1}{\ell}\left(n+\frac{1}{2} \right)\left[-0.19+0.05 (\alpha-\alpha_c) \right] \simeq -\frac{0.19}{\ell}\left(n+\frac{1}{2} \right)+ 1.63 \left(n+\frac{1}{2} \right) \ell \ T_\text{H}^2.
\end{align}
\end{subequations}

\section{Conclusions}
\label{Sec:Conclusions}
In this paper, we explored the possibility of   relating   long-range quantum gravity corrections 	at horizon scales with  the absence of the  central singularity in the Schwarzschild black hole. In order to achieve this, we explicitly constructed a general class of static, non-singular, asymptotically-flat black-hole models as exact solutions of Einstein's field equations sourced by an anisotropic fluid  with EoS $p_\parallel=-\varepsilon$. A natural physical consequence of this choice is the fact that these models possess a dS core. This causes a violation of the strong energy condition near the black-hole center, which therefore allows to circumvent Penrose's singularity theorem.

Analogously to what has been done   at   galactic and cosmological level  \cite{Cadoni:2017evg, Cadoni:2018dnd,Tuveri:2019zor, Cadoni:2020izk, Cadoni:2020jxe, Cadoni:2021zsl}, the anisotropic-fluid source is used to give a multi-scale description of the   gravitational system and an effective description  of the quantum black hole. In our models, quantum corrections are effectively encoded in a \textit{single} additional scale $\ell$, which appears as an extra "quantum hair" for the black hole and is related to the dS length $\hat L$, smearing the would-be central singularity of the Schwarzschild black hole. 

Apart from solving the singularity problem and allowing for quantum effects at horizon scales, our general black-hole solutions have also other rather interesting geometric and thermodynamic features. They allow for the presence of two horizons, an outer event horizon and an inner Cauchy one. When these two merge in a single horizon, we have an extremal configuration, which is a zero-temperature, zero-entropy state, whose near-horizon geometry factors as AdS$_2 \times$ S$^2$. This fact could be very useful in addressing the black-hole information puzzle.

The previous features determine a rather non-trivial thermodynamic behavior. The presence of the quantum hair $\ell$, not related to a thermodynamic potential, required a generalization of the area law for the black-hole entropy. We also found a metastable behavior and a phase transition, in which black holes closer to extremality, i.e. with $\ell \sim \RS$, are always thermodynamically preferred with respect to those with $\ell \gg \RS$. This means that our "quantum  black holes" are thermodynamically  preferred with respect to those  in which quantum corrections are irrelevant. This provided further evidence for the relevance of quantum effects at horizon scales.

 For $\ell \gtrsim \RS$, instead, we found that our models represent horizonless compact object, which however were not investigated in depth in the present paper and will be matter for further investigations. \\

Finally, we have also shown that, when $\ell\ll \RS$,  the phenomenology of our non-singular black-hole solutions is almost indistinguishable from the standard Schwarzschild-solution one for an outside observer. On the other hand, for $\ell \sim \RS$,  there could be manifest signatures of deviations from the standard behavior, both in the photon orbits and in the QNMs spectrum. 

For what concerns photon orbits, we have found stable bounded orbits, which are, however, internal to  (in the two-horizon model) or coinciding with  (in the extremal case) the horizon.  An interesting result is that for some horizonless models, unstable photon orbits do not form. 

On the QNMs side, we have investigated the spectrum of quasi-normal frequencies in the eikonal limit and in the near-extremal limit. We found a scaling behavior with the temperature and a dependence from the quantum hair $\ell$. These features have a manifest experimental signature, which could be experimentally accessible in the near future by the next  generation of GW detectors. Taking a near-horizon limit together with a near-extremal one, instead, we found an imaginary part which vanishes with the temperature. Our findings only partially confirm Hod's conjecture on zero-damped QNMs for standard extremal black holes \cite{Hod:2008se,Hod:2008zz,Hod:2011zzd,Hod:2012zzb,Hod:2015hga,Zimmerman:2015trm,Joykutty:2021fgj}.  On the other hand,  this result is perfectly consistent with  the AdS$_2 \times$ S$^2$ behavior of the near-horizon, near-extremal metric and with what is known about the QNMs spectrum of 2D AdS gravity models \cite{LopezOrtega:2011np, Cordero:2012je, Kettner:2004aw, Bhattacharjee:2020nul, Cadoni:2021qfn}.

We have confirmed our general results by thoroughly investigating two particular cases, which represent two widely-known non-singular black-hole models, namely the Hayward and the gaussian-core metrics. We have revisited these models considering the quantum deformation parameter of the same order of magnitude of the Schwarzschild radius $\RS$. This perspective is completely different from the standard approach to these models, where quantum effects, and hence $\ell$, are of the order of magnitude of Planck length, implying extremely small and phenomenologically irrelevant deviations from standard behavior.

\section*{Acknowledgements} 
We thank S. Liberati for useful discussions.  

\begin{appendix}
\section{Geodesic completness of the spacetime}
\label{appendix}
In this appendix we show that space-times described by \cref{f10} are geodesically complete. We start from Raychaudhuri's equation, which describes the evolution of a  time-like geodesic congruence $\Theta$
\begin{equation}
\frac{d\Theta}{d\tau}= -\frac{1}{3}\Theta^2 -\sigma^{\mu\nu}\sigma_{\mu\nu}+\omega^{\mu\nu}\omega_{\mu\nu}-R_{\mu\nu}u^{\mu}u^{\nu},
\end{equation}
where $\tau$ is the proper time, $u^{\mu}= dx^\mu/d\tau$ the proper time-like velocity, while $\sigma_{\mu\nu}= \Theta_{\mu\nu}-\frac{1}{3}\Theta h_{\mu\nu}$ is the shear tensor ($h_{\mu\nu}=g_{\mu\nu}+u_\mu u_\nu$ is the transverse metric) and $\omega_{\mu\nu}=h^\alpha_\mu h^\beta_\nu u_{[\alpha;\beta]}$ is the vorticity tensor. 
If we consider geodesics to be hypersurface orthogonal, then $\omega_{\mu\nu}=0$. Since both the shear and the vorticity tensors are purely spatial, i.e. $\omega_{\mu\nu}\omega^{\mu \nu}\geq 0$, $\sigma_{\mu\nu}\sigma^{\mu\nu}\geq 0$, and if we assume \textit{the strong energy condition to hold}, i.e. $R_{\mu\nu}u^\mu u^\nu \geq 0$, we expect in all generality a focusing of the geodesic congruence, i.e. $d\Theta/d\tau \leq 0$ and the formation of caustics, which represent  singularities of the congruence.  This is the essence of the original Penrose singularity theorem \cite{Penrose:1964wq}. 
Let us now show that caustics cannot form for the models described by \cref{f10}. We consider first a generic metric function $A$ and time-like geodesics
\begin{equation}
g_{\mu\nu}u^\mu u^\nu=-1,
\end{equation}
and  focus on geodesics on the plane $\theta=\text{constant}$ and $\varphi=\text{constant}$. We then have $u^\theta=u^\varphi =0$. We are left with
\begin{equation}
g_{\mu\nu}u^\mu u^\nu=-1 \Rightarrow -A(r) (u^0)^2+\frac{(u^r)^2}{A(r)}=-1.
\end{equation}
But $u^0 = dt/d\tau=1/A(r)$ (since the metric redshift factor determines the relation between the coordinate and proper time), and therefore
\begin{equation}
-\frac{1}{A(r)}+\frac{(u^r)^2}{A(r)}=-1 \Rightarrow u^r = \pm \sqrt{1-A(r)}.
\end{equation}
This yields  the  congruence 
\begin{equation}
\label{fgh1}
\Theta = \frac{1}{r^2}\partial_r\left(r^2 u^r \right) = \frac{1}{r^2}\left[2 r u^r + r^2 \frac{du^r}{dr}\right]=\pm \left( \frac{2}{r} \sqrt{1-A(r)}-\frac{A'(r)}{2\sqrt{1-A(r)}}\right).
\end{equation}
The geodesic congruence evolution as a  function of the  proper time can therefore be written as
\begin{equation}\label{geodesicevolutiongeneral}
\frac{d\Theta}{d\tau}=\frac{d\Theta}{dr}\frac{dr}{d\tau}= \Theta' u^r=-\frac{2(1-A(r))}{r^2}-\frac{A'(r)}{r}-\frac{A''(r)}{2}-\frac{A'(r)^2}{4(1-A(r))},
\end{equation}
where we used the fact that $\Theta$ is a function of $r$ only.

Let us now specialize \cref{geodesicevolutiongeneral} to the case given by \cref{f10}. We get
\begin{equation}\label{geodesicevolutiony}
\frac{d\Theta}{d\tau}=-\frac{2\alpha F(y)}{y^2 \ell^2}+\frac{\alpha}{y \ell^2}F'(y) +\frac{\alpha}{2\ell^2} F''(y)-\frac{\alpha}{4\ell^2}\frac{F'(y)^2}{F(y)}
\end{equation}
Since, for large $y$, our general model behaves essentially as the Schwarzschild black hole, we focus on the behavior of the geodesics bundle in the core of the object, i.e. for $y \to 0$, where $F \sim y^2$ according to \cref{hhhh}. Therefore, $F' \sim 2 y$ and $F'' \sim 2$, and thus
\begin{equation}
\frac{d\theta}{d\tau}\sim -\frac{2\alpha}{\ell^2}+\frac{2\alpha}{\ell^2}+\frac{\alpha}{\ell^2}-\frac{\alpha}{\ell^2}\sim 0,
\end{equation}
so that we do not encounter a caustic in the center, the bundle of geodesics is defocused and therefore  they can be extended beyond $r=0$. This is consistent with the form of the  Penrose diagram for such models (see e.g. Refs. \cite{Ansoldi:2008jw,Hayward:2005gi}), which  shows that, apart from the presence of the central singularity,  the maximal extension of these space-times is similar to  that of RN black holes.

These results can be further confirmed by computing $d\Theta/d\tau$ for the two specific models investigated in detail in the paper, namely the Hayward and the gaussian-core black holes.

For the Hayward  black hole, plugging the function $F$ \eqref{haywbh} into \cref{geodesicevolutiony} yields
\begin{equation}
\frac{d\Theta}{d\tau}= -\frac{9\alpha y^3\left(y^3+4 \right)}{4 \ell^2\left(1+y^3 \right)^3}.
\end{equation}
Near $y \sim 0$, we have $\frac{d\Theta}{d\tau}\sim -\frac{9\alpha y^3}{\ell^2}+\mathcal{O}(y^4)$, so no caustic is present in the interior, at $r=0$.

In the gaussian-core black-hole case, using the function $F$ \eqref{metricgaussian} into \cref{geodesicevolutiony}, we get
\begin{equation}
\frac{d\Theta}{d\tau}=-\frac{\alpha e^{-2y^2}}{4\ell^2 \sqrt{\pi}y^3} \cdot \frac{9 \pi e^{2y^2} + 16 y^6+8\sqrt{\pi}e^{y^2}y^3 \left(2y^2-3 \right)-4e^{y^2}\left(9\sqrt{\pi}e^{y^2}-12y^3+8y^5 \right)\Gamma\left(\frac{3}{2},y^2 \right)+36 e^{2y^2}\Gamma^2\left(\frac{3}{2},y^2 \right)}{\sqrt{\pi}-2 \Gamma \left(\frac{3}{2}, y^2 \right)}.
\end{equation}
Near $y \sim 0$, we have $\frac{d\Theta}{d\tau} \sim -\frac{4\alpha y^2}{\sqrt{\pi}\ell^2}+\mathcal{O}(y^4)$, so that again caustics do not form and the space-time is geodesically complete.
\end{appendix}

\bibliography{NonSingularBH_PRD_Revised}

\begin{thebibliography}{100}

\bibitem{Penrose:1964wq}
R.~Penrose, ``{Gravitational collapse and space-time singularities},'' {\em
  \href{http://dx.doi.org/10.1103/PhysRevLett.14.57} {Phys. Rev. Lett.
  \textbf{14} (1965), 57-59}}.

\bibitem{Hawking:1970zqf}
S.~W. Hawking and R.~Penrose, ``{The Singularities of gravitational collapse
  and cosmology},'' {\em \href{http://dx.doi.org/10.1098/rspa.1970.0021} {Proc.
  Roy. Soc. Lond. A \textbf{314} (1970), 529-548}}.

\bibitem{Penrose:1969pc}
R.~Penrose, ``{Gravitational collapse: The role of general relativity},'' {\em
  \href{http://dx.doi.org/10.1023/A:1016578408204} {Riv. Nuovo Cim. \textbf{1}
  (1969), 252-276}}.

\bibitem{Hawking:1974sw}
S.~W. {Hawking}, ``{Particle creation by black holes},'' {\em
  \href{http://dx.doi.org/10.1007/BF02345020} {Commun. Math. Phys. \textbf{43}
  (1975), 199-220 [erratum: Commun. Math. Phys. \textbf{46} (1976), 206]}}.

\bibitem{Carballo-Rubio:2018jzw}
R.~Carballo-Rubio, F.~Di~Filippo, S.~Liberati, and M.~Visser,
  ``{Phenomenological aspects of black holes beyond general relativity},'' {\em
  \href{http://dx.doi.org/10.1103/PhysRevD.98.124009} {Phys. Rev. D \textbf{98}
  (2018) no.12, 124009}}.

\bibitem{Carballo-Rubio:2019fnb}
R.~Carballo-Rubio, F.~Di~Filippo, S.~Liberati, and M.~Visser, ``{Geodesically
  complete black holes},'' {\em
  \href{http://dx.doi.org/10.1103/PhysRevD.101.084047} {Phys. Rev. D
  \textbf{101} (2020), 084047}}.

\bibitem{Maeda:2021jdc}
H.~Maeda, ``{Quest for realistic non-singular black-hole geometries:
  Regular-center type},'' {\em
  \href{https://arxiv.org/abs/2107.04791}{{\ttfamily arXiv:2107.04791
  [gr-qc]}}}.

\bibitem{Simpson:2019mud}
A.~Simpson and M.~Visser, ``{Regular black holes with asymptotically Minkowski
  cores},'' {\em \href{http://dx.doi.org/10.3390/universe6010008} {Universe
  \textbf{6} (2019) no.1, 8}}.

\bibitem{Lobo:2020ffi}
F.~S.~N. Lobo, M.~E. Rodrigues, M.~V. d.~S. Silva, A.~Simpson, and M.~Visser,
  ``{Novel black-bounce spacetimes: wormholes, regularity, energy conditions,
  and causal structure},'' {\em
  \href{http://dx.doi.org/10.1103/PhysRevD.103.084052} {Phys. Rev. D
  \textbf{103} (2021) no.8, 084052}}.

\bibitem{Mazza:2021rgq}
J.~Mazza, E.~Franzin, and S.~Liberati, ``{A novel family of rotating black hole
  mimickers},'' {\em \href{http://dx.doi.org/10.1088/1475-7516/2021/04/082}
  {JCAP \textbf{04} (2021), 082}}.

\bibitem{Franzin:2021vnj}
E.~Franzin, S.~Liberati, J.~Mazza, A.~Simpson, and M.~Visser, ``{Charged
  black-bounce spacetimes},'' {\em
  \href{http://dx.doi.org/10.1088/1475-7516/2021/07/036} {JCAP \textbf{07}
  (2021), 036}}.

\bibitem{Ayon-Beato:1999qin}
E.~Ayon-Beato and A.~Garcia, ``{Nonsingular charged black hole solution for
  nonlinear source},'' {\em \href{http://dx.doi.org/10.1023/A:1026640911319}
  {Gen. Rel. Grav. \textbf{31} (1999), 629-633}}.

\bibitem{Bronnikov:2000vy}
K.~A. Bronnikov, ``{Regular magnetic black holes and monopoles from nonlinear
  electrodynamics},'' {\em \href{http://dx.doi.org/10.1103/PhysRevD.63.044005}
  {Phys. Rev. D \textbf{63} (2001), 044005}}.

\bibitem{Dymnikova:2004zc}
I.~Dymnikova, ``{Regular electrically charged structures in nonlinear
  electrodynamics coupled to general relativity},'' {\em
  \href{http://dx.doi.org/10.1088/0264-9381/21/18/009} {Class. Quant. Grav.
  \textbf{21} (2004), 4417-4429}}.

\bibitem{Culetu:2014lca}
H.~Culetu, ``{On a regular charged black hole with a nonlinear electric
  source},'' {\em \href{http://dx.doi.org/10.1007/s10773-015-2521-6} {Int. J.
  Theor. Phys. \textbf{54} (2015) no.8, 2855-2863}}.

\bibitem{Banerjee:2022len}
I.~Banerjee, ``{Signatures of regular black holes from the quasar continuum
  spectrum},'' {\em \href{https://arxiv.org/abs/2206.06899}{{\ttfamily
  arXiv:2206.06899 [gr-qc]}}}.

\bibitem{Bokulic:2022cyk}
A.~Bokuli\'c, T.~Juri\'c, and I.~Smoli\'c, ``{Constraints on singularity
  resolution by nonlinear electrodynamics},'' {\em
  \href{https://arxiv.org/abs/2206.07064}{{\ttfamily arXiv:2206.07064
  [gr-qc]}}}.

\bibitem{Tseytlin:1995uq}
A.~A. Tseytlin, ``{On singularities of spherically symmetric backgrounds in
  string theory},'' {\em \href{http://dx.doi.org/10.1016/0370-2693(95)01228-7}
  {Phys. Lett. B \textbf{363} (1995), 223-229}}.

\bibitem{Lawrence:1995ct}
A.~E. Lawrence and E.~J. Martinec, ``{String field theory in curved space-time
  and the resolution of space - like singularities},'' {\em
  \href{http://dx.doi.org/10.1088/0264-9381/13/1/007} {Class. Quant. Grav.
  \textbf{13} (1996), 63-96}}.

\bibitem{Horowitz:1989bv}
G.~T. Horowitz and A.~R. Steif, ``{Space-Time Singularities in String
  Theory},'' {\em \href{http://dx.doi.org/10.1103/PhysRevLett.64.260} {Phys.
  Rev. Lett. \textbf{64} (1990), 260}}.

\bibitem{Modesto:2004xx}
L.~Modesto, ``{Disappearance of black hole singularity in quantum gravity},''
  {\em \href{http://dx.doi.org/10.1103/PhysRevD.70.124009} {Phys. Rev. D
  \textbf{70} (2004), 124009}}.

\bibitem{Nicolini:2005vd}
P.~Nicolini, A.~Smailagic, and E.~Spallucci, ``{Noncommutative geometry
  inspired Schwarzschild black hole},'' {\em
  \href{http://dx.doi.org/10.1016/j.physletb.2005.11.004} {Phys. Lett. B
  \textbf{632} (2006), 547-551}}.

\bibitem{Modesto:2008im}
L.~Modesto, ``{Semiclassical loop quantum black hole},'' {\em
  \href{http://dx.doi.org/10.1007/s10773-010-0346-x} {Int. J. Theor. Phys.
  \textbf{49} (2010), 1649-1683}}.

\bibitem{Nicolini:2008aj}
P.~Nicolini, ``{Noncommutative Black Holes, The Final Appeal To Quantum
  Gravity: A Review},'' {\em \href{http://dx.doi.org/10.1142/S0217751X09043353}
  {Int. J. Mod. Phys. A \textbf{24} (2009), 1229-1308}}.

\bibitem{Hossenfelder:2009fc}
S.~Hossenfelder, L.~Modesto, and I.~Premont-Schwarz, ``{A Model for
  non-singular black hole collapse and evaporation},'' {\em
  \href{http://dx.doi.org/10.1103/PhysRevD.81.044036} {Phys. Rev. D \textbf{81}
  (2010), 044036}}.

\bibitem{Modesto:2010uh}
L.~Modesto, J.~W. Moffat, and P.~Nicolini, ``{Black holes in an ultraviolet
  complete quantum gravity},'' {\em
  \href{http://dx.doi.org/10.1016/j.physletb.2010.11.046} {Phys. Lett. B
  \textbf{695} (2011), 397-400}}.

\bibitem{Spallucci:2011rn}
E.~Spallucci and S.~Ansoldi, ``{Regular black holes in UV self-complete quantum
  gravity},'' {\em \href{http://dx.doi.org/10.1016/j.physletb.2011.06.005}
  {Phys. Lett. B \textbf{701} (2011), 471-474}}.

\bibitem{Sprenger:2012uc}
M.~Sprenger, P.~Nicolini, and M.~Bleicher, ``{Physics on Smallest Scales - An
  Introduction to Minimal Length Phenomenology},'' {\em
  \href{http://dx.doi.org/10.1088/0143-0807/33/4/853} {Eur. J. Phys.
  \textbf{33} (2012), 853-862}}.

\bibitem{Bambi:2013caa}
C.~Bambi, D.~Malafarina, and L.~Modesto, ``{Non-singular quantum-inspired
  gravitational collapse},'' {\em
  \href{http://dx.doi.org/10.1103/PhysRevD.88.044009} {Phys. Rev. D \textbf{88}
  (2013), 044009}}.

\bibitem{Frolov:2014jva}
V.~P. Frolov, ``{Information loss problem and a 'black hole` model with a
  closed apparent horizon},'' {\em
  \href{http://dx.doi.org/10.1007/JHEP05(2014)049} {JHEP \textbf{05} (2014),
  049}}.

\bibitem{Casadio:2014pia}
R.~Casadio, O.~Micu, and P.~Nicolini, ``{Minimum length effects in black hole
  physics},'' {\em \href{http://dx.doi.org/10.1007/978-3-319-10852-0_10}
  {Fundam. Theor. Phys. \textbf{178} (2015), 293-322}}.

\bibitem{Binetti:2022xdi}
E.~Binetti, M.~Del~Piano, S.~Hohenegger, F.~Pezzella, and F.~Sannino, ``{The
  Effective Theory of Quantum Black Holes},'' {\em
  \href{https://arxiv.org/abs/2203.13515}{{\ttfamily arXiv:2203.13515
  [gr-qc]}}}.

\bibitem{Almheiri:2012rt}
A.~Almheiri, D.~Marolf, J.~Polchinski, and J.~Sully, ``{Black Holes:
  Complementarity or Firewalls?},'' {\em
  \href{http://dx.doi.org/10.1007/JHEP02(2013)062} {JHEP \textbf{02} (2013),
  062}}.

\bibitem{Penington:2019kki}
G.~{Penington}, S.~H. {Shenker}, D.~{Stanford}, and Z.~{Yang}, ``{Replica
  wormholes and the black hole interior},'' {\em
  \href{https://arxiv.org/abs/1911.11977}{{\ttfamily arXiv:1911.11977
  [hep-th]}}}.

\bibitem{Almheiri:2019qdq}
A.~Almheiri, T.~Hartman, J.~Maldacena, E.~Shaghoulian, and A.~Tajdini,
  ``{Replica Wormholes and the Entropy of Hawking Radiation},'' {\em
  \href{http://dx.doi.org/10.1007/JHEP05(2020)013} {JHEP \textbf{05} (2020),
  013}}.

\bibitem{Almheiri:2020cfm}
A.~Almheiri, T.~Hartman, J.~Maldacena, E.~Shaghoulian, and A.~Tajdini, ``{The
  entropy of Hawking radiation},'' {\em
  \href{http://dx.doi.org/10.1103/RevModPhys.93.035002} {Rev. Mod. Phys.
  \textbf{93} (2021) no.3, 035002}}.

\bibitem{Bousso:2022ntt}
R.~Bousso, X.~Dong, N.~Engelhardt, T.~Faulkner, T.~Hartman, S.~H. Shenker, and
  D.~Stanford, ``{Snowmass White Paper: Quantum Aspects of Black Holes and the
  Emergence of Spacetime},'' {\em
  \href{https://arxiv.org/abs/2201.03096}{{\ttfamily arXiv:2201.03096
  [hep-th]}}}.

\bibitem{Giddings:2012gc}
S.~B. Giddings, ``{Nonviolent nonlocality},'' {\em
  \href{http://dx.doi.org/10.1103/PhysRevD.88.064023} {Phys. Rev. D \textbf{88}
  (2013), 064023}}.

\bibitem{Giddings:2021qas}
S.~B. Giddings, ``{A ''black hole theorem,'' and its implications},'' {\em
  \href{https://arxiv.org/abs/2110.10690}{{\ttfamily arXiv:2110.10690
  [hep-th]}}}.

\bibitem{Giddings:2022jda}
S.~B. Giddings, ``{The deepest problem: some perspectives on quantum
  gravity},'' {\em \href{https://arxiv.org/abs/2202.08292}{{\ttfamily
  arXiv:2202.08292 [hep-th]}}}.

\bibitem{Mathur:2005zp}
S.~D. Mathur, ``{The Fuzzball proposal for black holes: An Elementary
  review},'' {\em \href{http://dx.doi.org/10.1002/prop.200410203} {Fortsch.
  Phys. \textbf{53} (2005), 793-827}}.

\bibitem{Mathur:2019dhf}
S.~D. Mathur, ``{The nature of the gravitational vacuum},'' {\em
  \href{http://dx.doi.org/10.1142/S021827181944005X} {Int. J. Mod. Phys. D
  \textbf{28} (2019) no.14, 1944005}}.

\bibitem{Mathur:2020ely}
S.~D. Mathur, ``{The VECRO hypothesis},'' {\em
  \href{http://dx.doi.org/10.1142/S0218271820300098} {Int. J. Mod. Phys. D
  \textbf{29} (2020) no.15, 2030009}}.

\bibitem{Verlinde:2016toy}
E.~P. Verlinde, ``{Emergent Gravity and the Dark Universe},'' {\em
  \href{http://dx.doi.org/10.21468/SciPostPhys.2.3.016} {SciPost Phys.
  \textbf{2} (2017) no.3, 016}}.

\bibitem{Dvali:2011aa}
G.~Dvali and C.~Gomez, ``{Black Hole's Quantum N-Portrait},'' {\em
  \href{http://dx.doi.org/10.1002/prop.201300001} {Fortsch. Phys. \textbf{61}
  (2013), 742-767}}.

\bibitem{Dvali:2013eja}
G.~Dvali and C.~Gomez, ``{Quantum Compositeness of Gravity: Black Holes, AdS
  and Inflation},'' {\em \href{http://dx.doi.org/10.1088/1475-7516/2014/01/023}
  {JCAP \textbf{01} (2014), 023}}.

\bibitem{Dvali:2020wqi}
G.~Dvali, ``{Entropy Bound and Unitarity of Scattering Amplitudes},'' {\em
  \href{http://dx.doi.org/10.1007/JHEP03(2021)126} {JHEP \textbf{03} (2021),
  126}}.

\bibitem{Casadio:2015lis}
R.~Casadio, A.~Giugno, O.~Micu, and A.~Orlandi, ``{Thermal BEC black holes},''
  {\em \href{http://dx.doi.org/10.3390/e17106893} {Entropy \textbf{17} (2015),
  6893-6924}}.

\bibitem{Casadio:2016zpl}
R.~Casadio, A.~Giugno, and A.~Giusti, ``{Matter and gravitons in the
  gravitational collapse},'' {\em
  \href{http://dx.doi.org/10.1016/j.physletb.2016.10.058} {Phys. Lett. B
  \textbf{763} (2016), 337-340}}.

\bibitem{Cadoni:2020mgb}
M.~Cadoni, M.~Tuveri, and A.~P. Sanna, ``{Long-Range Quantum Gravity},'' {\em
  \href{http://dx.doi.org/10.3390/sym12091396} {Symmetry \textbf{12} (2020)
  no.9, 1396}}.

\bibitem{Dvali:2010bf}
G.~Dvali and C.~Gomez, ``{Self-Completeness of Einstein Gravity},'' {\em
  \href{https://arxiv.org/abs/1005.3497}{{\ttfamily arXiv:1005.3497
  [hep-th]}}}.

\bibitem{Dvali:2010jz}
G.~Dvali, G.~F. Giudice, C.~Gomez, and A.~Kehagias, ``{UV-Completion by
  Classicalization},'' {\em \href{http://dx.doi.org/10.1007/JHEP08(2011)108}
  {JHEP \textbf{08} (2011), 108}}.

\bibitem{Dvali:2011th}
G.~Dvali, C.~Gomez, and A.~Kehagias, ``{Classicalization of Gravitons and
  Goldstones},'' {\em \href{http://dx.doi.org/10.1007/JHEP11(2011)070} {JHEP
  \textbf{11} (2011), 070}}.

\bibitem{Maggiore:2007nq}
M.~Maggiore, ``{The Physical interpretation of the spectrum of black hole
  quasinormal modes},'' {\em
  \href{http://dx.doi.org/10.1103/PhysRevLett.100.141301} {Phys. Rev. Lett.
  \textbf{100} (2008), 141301}}.

\bibitem{Cadoni:2021jer}
M.~Cadoni, M.~Oi, and A.~P. Sanna, ``{Quasinormal modes and microscopic
  structure of the Schwarzschild black hole},'' {\em
  \href{http://dx.doi.org/10.1103/PhysRevD.104.L121502} {Phys. Rev. D
  \textbf{104} (2021) no.12, L121502}}.

\bibitem{Cadoni:2018dnd}
M.~Cadoni, R.~Casadio, A.~Giusti, and M.~Tuveri, ``{Emergence of a Dark Force
  in Corpuscular Gravity},'' {\em
  \href{http://dx.doi.org/10.1103/PhysRevD.97.044047} {Phys. Rev. D \textbf{97}
  (2018) no.4, 044047}}.

\bibitem{Cadoni:2017evg}
M.~Cadoni, R.~Casadio, A.~Giusti, W.~M\"uck, and M.~Tuveri, ``{Effective Fluid
  Description of the Dark Universe},'' {\em
  \href{http://dx.doi.org/10.1016/j.physletb.2017.11.058} {Phys. Lett. B
  \textbf{776} (2018), 242-248}}.

\bibitem{Tuveri:2019zor}
M.~Tuveri and M.~Cadoni, ``{Galactic dynamics and long-range quantum
  gravity},'' {\em \href{http://dx.doi.org/10.1103/PhysRevD.100.024029} {Phys.
  Rev. D \textbf{100} (2019) no.2, 024029}}.

\bibitem{Cadoni:2020izk}
M.~Cadoni, A.~P. Sanna, and M.~Tuveri, ``{Anisotropic fluid cosmology: An
  alternative to dark matter?},'' {\em
  \href{http://dx.doi.org/10.1103/PhysRevD.102.023514} {Phys. Rev. D
  \textbf{102} (2020) no.2, 023514}}.

\bibitem{Cadoni:2020jxe}
M.~Cadoni and A.~P. Sanna, ``{Emergence of a cosmological constant in
  anisotropic fluid cosmology},'' {\em
  \href{http://dx.doi.org/10.1142/S0217751X21501566} {Int. J. Mod. Phys. A
  \textbf{36} (2021) no.21, 2150156}}.

\bibitem{Cadoni:2021zsl}
M.~Cadoni and A.~P. Sanna, ``{Unified description of galactic dynamics and the
  cosmological constant},'' {\em
  \href{http://dx.doi.org/10.1088/1361-6382/abfd92} {Class. Quant. Grav.
  \textbf{38} (2021) no.13, 135004}}.

\bibitem{Oshita:2019sat}
N.~Oshita, Q.~Wang, and N.~Afshordi, ``{On Reflectivity of Quantum Black Hole
  Horizons},'' {\em \href{http://dx.doi.org/10.1088/1475-7516/2020/04/016}
  {JCAP \textbf{04} (2020), 016}}.

\bibitem{Wang:2019rcf}
Q.~Wang, N.~Oshita, and N.~Afshordi, ``{Echoes from Quantum Black Holes},''
  {\em \href{http://dx.doi.org/10.1103/PhysRevD.101.024031} {Phys. Rev. D
  \textbf{101} (2020) no.2, 024031}}.

\bibitem{Chakraborty:2022zlq}
S.~Chakraborty, E.~Maggio, A.~Mazumdar, and P.~Pani, ``{Implications of the
  quantum nature of the black hole horizon on the gravitational-wave
  ringdown},'' {\em \href{https://arxiv.org/abs/2202.09111}{{\ttfamily
  arXiv:2202.09111 [gr-qc]}}}.

\bibitem{Hawking:1976ra}
S.~W. Hawking, ``{Breakdown of Predictability in Gravitational Collapse},''
  {\em \href{http://dx.doi.org/10.1103/PhysRevD.14.2460} {Phys. Rev. D
  \textbf{14} (1976), 2460-2473}}.

\bibitem{Page:1993wv}
D.~N. Page, ``{Information in black hole radiation},'' {\em
  \href{http://dx.doi.org/10.1103/PhysRevLett.71.3743} {Phys. Rev. Lett.
  \textbf{71} (1993), 3743-3746}}.

\bibitem{Mathur:2009hf}
S.~D. Mathur, ``{The Information paradox: A Pedagogical introduction},'' {\em
  \href{http://dx.doi.org/10.1088/0264-9381/26/22/224001} {Class. Quant. Grav.
  \textbf{26} (2009), 224001}}.

\bibitem{Kitaev:2017awl}
A.~Kitaev and S.~J. Suh, ``{The soft mode in the Sachdev-Ye-Kitaev model and
  its gravity dual},'' {\em
  \href{http://dx.doi.org/10.1007/JHEP05(2018)183}{JHEP \textbf{05} (2018),
  183}}.

\bibitem{Almheiri:2019hni}
A.~Almheiri, R.~Mahajan, J.~Maldacena, and Y.~Zhao, ``{The Page curve of
  Hawking radiation from semiclassical geometry},'' {\em
  \href{http://dx.doi.org/10.1007/JHEP03(2020)149} {JHEP \textbf{03} (2020),
  149}}.

\bibitem{LopezOrtega:2011np}
A.~Lopez-Ortega, ``{Entropy spectra of single horizon black holes in two
  dimensions},'' {\em \href{http://dx.doi.org/10.1142/S0218271811020524} {Int.
  J. Mod. Phys. D \textbf{20} (2011), 2525-2542}}.

\bibitem{Cordero:2012je}
R.~Cordero, A.~Lopez-Ortega, and I.~Vega-Acevedo, ``{Quasinormal frequencies of
  asymptotically anti-de Sitter black holes in two dimensions},'' {\em
  \href{http://dx.doi.org/10.1007/s10714-011-1316-1} {Gen. Rel. Grav.
  \textbf{44} (2012), 917-940}}.

\bibitem{Kettner:2004aw}
J.~Kettner, G.~Kunstatter, and A.~J.~M. Medved, ``{Quasinormal modes for single
  horizon black holes in generic 2-d dilaton gravity},'' {\em
  \href{http://dx.doi.org/10.1088/0264-9381/21/23/002} {Class. Quant. Grav.
  \textbf{21} (2004), 5317-5332}}.

\bibitem{Bhattacharjee:2020nul}
S.~Bhattacharjee, S.~Sarkar, and A.~Bhattacharyya, ``{Scalar perturbations of
  black holes in Jackiw-Teitelboim gravity},'' {\em
  \href{http://dx.doi.org/10.1103/PhysRevD.103.024008} {Phys. Rev. D
  \textbf{103} (2021) no.2, 024008}}.

\bibitem{Cadoni:2021qfn}
M.~Cadoni, M.~Oi, and A.~P. Sanna, ``{Quasi-normal modes and microscopic
  description of 2D black holes},'' {\em
  \href{http://dx.doi.org/10.1007/JHEP01(2022)087} {JHEP \textbf{01} (2022),
  087}}.

\bibitem{Hod:2008se}
S.~Hod, ``{Quasinormal resonances of near-extremal Kerr-Newman black holes},''
  {\em \href{http://dx.doi.org/10.1016/j.physletb.2008.08.002} {Phys. Lett. B
  \textbf{666} (2008), 483-485}}.

\bibitem{Hod:2008zz}
S.~Hod, ``{Slow relaxation of rapidly rotating black holes},'' {\em
  \href{http://dx.doi.org/10.1103/PhysRevD.78.084035} {Phys. Rev. D \textbf{78}
  (2008), 084035}}.

\bibitem{Hod:2011zzd}
S.~Hod, ``{Quasinormal resonances of a massive scalar field in a near-extremal
  Kerr black hole spacetime},'' {\em
  \href{http://dx.doi.org/10.1016/10.1103/PhysRevD.84.044046} {Phys. Rev. D
  \textbf{84} (2011), 044046}}.

\bibitem{Hod:2012zzb}
S.~Hod, ``{Quasinormal resonances of a charged scalar field in a charged
  Reissner-Nordstroem black-hole spacetime: A WKB analysis},'' {\em
  \href{http://dx.doi.org/10.1016/j.physletb.2012.03.010} {Phys. Lett. B
  \textbf{710} (2012), 349-351}}.

\bibitem{Hod:2015hga}
S.~Hod, ``{Universality in the relaxation dynamics of the composed
  black-hole-charged-massive-scalar-field system: The role of quantum Schwinger
  discharge},'' {\em \href{http://dx.doi.org/10.1016/j.physletb.2015.06.019}
  {Phys. Lett. B \textbf{747} (2015), 339-344}}.

\bibitem{Zimmerman:2015trm}
A.~Zimmerman and Z.~Mark, ``{Damped and zero-damped quasinormal modes of
  charged, nearly extremal black holes},'' {\em
  \href{http://dx.doi.org/10.1103/PhysRevD.93.044033} {Phys. Rev. D \textbf{93}
  (2016) no.4, 044033 [erratum: Phys. Rev. D \textbf{93} (2016) no.8,
  089905]}}.

\bibitem{Joykutty:2021fgj}
J.~Joykutty, ``{Existence of Zero-damped Quasinormal Frequencies for Nearly
  Extremal Black Holes},'' {\em
  \href{https://arxiv.org/abs/2112.05669}{{\ttfamily arXiv:2112.05669
  [gr-qc]}}}.

\bibitem{Ansoldi:2006vg}
S.~Ansoldi, P.~Nicolini, A.~Smailagic, and E.~Spallucci, ``{Noncommutative
  geometry inspired charged black holes},'' {\em
  \href{http://dx.doi.org/10.1016/j.physletb.2006.12.020} {Phys. Lett. B
  \textbf{645} (2007), 261-266}}.

\bibitem{Modesto:2010rv}
L.~Modesto and P.~Nicolini, ``{Charged rotating noncommutative black holes},''
  {\em \href{http://dx.doi.org/10.1103/PhysRevD.82.104035} {Phys. Rev. D
  \textbf{82} (2010), 104035}}.

\bibitem{Lan:2020fmn}
C.~Lan, Y.-G. Miao, and H.~Yang, ``{Quasinormal modes and phase transitions of
  regular black holes},'' {\em
  \href{http://dx.doi.org/10.1016/j.nuclphysb.2021.115539} {Nucl. Phys. B
  \textbf{971} (2021), 115539}}.

\bibitem{Cadoni:2006ww}
M.~Cadoni, ``{Conformal symmetry of gravity and the cosmological constant
  problem},'' {\em \href{http://dx.doi.org/10.1016/j.physletb.2006.10.009}
  {Phys. Lett. B \textbf{642} (2006), 525-529}}.

\bibitem{Bonanno:2000ep}
A.~Bonanno and M.~Reuter, ``{Renormalization group improved black hole
  space-times},'' {\em \href{http://dx.doi.org/10.1103/PhysRevD.62.043008}
  {Phys. Rev. D \textbf{62} (2000), 043008}}.

\bibitem{Niedermaier:2006wt}
M.~Niedermaier and M.~Reuter, ``{The Asymptotic Safety Scenario in Quantum
  Gravity},'' {\em \href{http://dx.doi.org/10.12942/lrr-2006-5} {Living Rev.
  Rel. \textbf{9} (2006), 5-173}}.

\bibitem{Bonanno:2020bil}
A.~Bonanno, A.~Eichhorn, H.~Gies, J.~M. Pawlowski, R.~Percacci, M.~Reuter,
  F.~Saueressig, and G.~P. Vacca, ``{Critical reflections on asymptotically
  safe gravity},'' {\em \href{http://dx.doi.org/10.3389/fphy.2020.00269}
  {Front. in Phys. \textbf{8} (2020), 269}}.

\bibitem{Adeifeoba:2018ydh}
A.~Adeifeoba, A.~Eichhorn, and A.~Platania, ``{Towards conditions for
  black-hole singularity-resolution in asymptotically safe quantum gravity},''
  {\em \href{http://dx.doi.org/10.1088/1361-6382/aae6ef} {Class. Quant. Grav.
  \textbf{35} (2018) no.22, 225007}}.

\bibitem{Borissova:2022jqj}
J.~N. Borissova, A.~Held, and N.~Afshordi, ``{Scale-Invariance at the Core of
  Quantum Black Holes},'' {\em
  \href{https://arxiv.org/abs/2203.02559}{{\ttfamily arXiv:2203.02559
  [gr-qc]}}}.

\bibitem{Bayin:1985cd}
S.~S. Bayin, ``{Anisotropic fluids and cosmology},'' {\em
  \href{http://dx.doi.org/10.1086/164056} {Astrophys. J. \textbf{303} (1986),
  101-110}}.

\bibitem{cosenza1981some}
M.~Cosenza, L.~Herrera, M.~Esculpi, and L.~Witten, ``Some models of anisotropic
  spheres in general relativity,'' {\em
  \href{http://dx.doi.org/10.1063/1.524742} {J. Math. Phys. \textbf{22}, 118
  (1981)}}.

\bibitem{DeBenedictis:2005vp}
A.~DeBenedictis, D.~Horvat, S.~Ilijic, S.~Kloster, and K.~S. Viswanathan,
  ``{Gravastar solutions with continuous pressures and equation of state},''
  {\em \href{http://dx.doi.org/10.1088/0264-9381/23/7/007} {Class. Quant. Grav.
  \textbf{23} (2006), 2303-2316}}.

\bibitem{Hayward:2005gi}
S.~A. Hayward, ``{Formation and evaporation of regular black holes},'' {\em
  \href{http://dx.doi.org/10.1103/PhysRevLett.96.031103} {Phys. Rev. Lett.
  \textbf{96} (2006), 031103}}.

\bibitem{Chirenti:2007mk}
C.~B. M.~H. Chirenti and L.~Rezzolla, ``{How to tell a gravastar from a black
  hole},'' {\em \href{http://dx.doi.org/10.1088/0264-9381/24/16/013} {Class.
  Quant. Grav. \textbf{24} (2007), 4191-4206}}.

\bibitem{Chan:2011ayt}
R.~Chan, M.~F.~A. da~Silva, and P.~Rocha, ``{Gravastars and Black Holes of
  Anisotropic Dark Energy},'' {\em
  \href{http://dx.doi.org/10.1007/s10714-011-1178-6} {Gen. Rel. Grav.
  \textbf{43} (2011), 2223-2235}}.

\bibitem{MartinMoruno:2011rm}
P.~Martin~Moruno, N.~Montelongo~Garcia, F.~S.~N. Lobo, and M.~Visser,
  ``{Generic thin-shell gravastars},'' {\em
  \href{http://dx.doi.org/10.1088/1475-7516/2012/03/034} {JCAP \textbf{03}
  (2012), 034}}.

\bibitem{Aluri:2012re}
P.~Aluri, S.~Panda, M.~Sharma, and S.~Thakur, ``{Anisotropic universe with
  anisotropic sources},'' {\em
  \href{http://dx.doi.org/10.1088/1475-7516/2013/12/003} {JCAP \textbf{12}
  (2013), 003}}.

\bibitem{Culetu:2013fsa}
H.~Culetu, ``{On a regular modified Schwarzschild spacetime},'' {\em
  \href{https://arxiv.org/abs/1305.5964}{{\ttfamily arXiv:1305.5964 [gr-qc]}}}.

\bibitem{Harko:2013wsa}
T.~Harko and F.~S.~N. Lobo, ``{Cosmological anisotropy from non-comoving dark
  matter and dark energy},'' {\em
  \href{http://dx.doi.org/10.1088/1475-7516/2013/07/036} {JCAP \textbf{07}
  (2013), 036}}.

\bibitem{Raposo:2018rjn}
G.~Raposo, P.~Pani, M.~Bezares, C.~Palenzuela, and V.~Cardoso, ``{Anisotropic
  stars as ultracompact objects in General Relativity},'' {\em
  \href{http://dx.doi.org/10.1103/PhysRevD.99.104072} {Phys. Rev. D \textbf{99}
  (2019) no.10, 104072}}.

\bibitem{Beltracchi:2018ait}
P.~Beltracchi and P.~Gondolo, ``{Formation of dark energy stars},'' {\em
  \href{http://dx.doi.org/10.1103/PhysRevD.99.044037} {Phys. Rev. D \textbf{99}
  (2019) no.4, 044037}}.

\bibitem{Kumar:2021oxa}
J.~Kumar and P.~Bharti, ``{The classification of interior solutions of
  anisotropic fluid configurations},'' {\em
  \href{https://arxiv.org/abs/2112.12518}{{\ttfamily arXiv:2112.12518
  [gr-qc]}}}.

\bibitem{Musco:2021sva}
I.~Musco and T.~Papanikolaou, ``{Primordial black hole formation for an
  anisotropic perfect fluid: initial conditions and estimation of the
  threshold},'' {\em \href{https://arxiv.org/abs/2110.05982}{{\ttfamily
  arXiv:2110.05982 [gr-qc]}}}.

\bibitem{Hawking:1973uf}
S.~W. Hawking and G.~F.~R. Ellis, {\em {The Large Scale Structure of
  Space-Time}}.
\newblock Cambridge Monographs on Mathematical Physics, Cambridge University
  Press, (2011).

\bibitem{Poisson:1989zz}
E.~Poisson and W.~Israel, ``{Inner-horizon instability and mass inflation in
  black holes},'' {\em \href{http://dx.doi.org/10.1103/PhysRevLett.63.1663}
  {Phys. Rev. Lett. \textbf{63} (1989), 1663-1666}}.

\bibitem{Ori:1991zz}
A.~Ori, ``{Inner structure of a charged black hole: An exact mass-inflation
  solution},'' {\em \href{http://dx.doi.org/10.1103/PhysRevLett.67.789} {Phys.
  Rev. Lett. \textbf{67} (1991), 789-792}}.

\bibitem{Markovic:1994gy}
D.~Markovic and E.~Poisson, ``{Classical stability and quantum instability of
  black hole Cauchy horizons},'' {\em
  \href{http://dx.doi.org/10.1103/PhysRevLett.74.1280} {Phys. Rev. Lett.
  \textbf{74} (1995), 1280-1283}}.

\bibitem{Carballo-Rubio:2018pmi}
R.~Carballo-Rubio, F.~Di~Filippo, S.~Liberati, C.~Pacilio, and M.~Visser, ``{On
  the viability of regular black holes},'' {\em
  \href{http://dx.doi.org/10.1007/JHEP07(2018)023} {JHEP \textbf{07} (2018),
  023}}.

\bibitem{Carballo-Rubio:2021bpr}
R.~Carballo-Rubio, F.~Di~Filippo, S.~Liberati, C.~Pacilio, and M.~Visser,
  ``{Inner horizon instability and the unstable cores of regular black
  holes},'' {\em \href{http://dx.doi.org/10.1007/JHEP05(2021)132} {JHEP
  \textbf{05} (2021), 132}}.

\bibitem{Bonanno:2020fgp}
A.~Bonanno, A.-P. Khosravi, and F.~Saueressig, ``{Regular black holes with
  stable cores},'' {\em \href{http://dx.doi.org/10.1103/PhysRevD.103.124027}
  {Phys. Rev. D \textbf{103} (2021) no.12, 124027}}.

\bibitem{Giugno:2017xtl}
A.~Giugno, A.~Giusti, and A.~Helou, ``{Horizon quantum fuzziness for
  non-singular black holes},'' {\em
  \href{http://dx.doi.org/10.1140/epjc/s10052-018-5715-2} {Eur. Phys. J. C
  \textbf{78} (2018) no.3, 208}}.

\bibitem{Casadio:2015qaq}
R.~Casadio, A.~Giugno, and O.~Micu, ``{Horizon quantum mechanics: A
  hitchhiker\textquoteright{}s guide to quantum black holes},'' {\em
  \href{http://dx.doi.org/10.1142/S0218271816300068} {Int. J. Mod. Phys. D
  \textbf{25} (2016) no.02, 1630006}}.

\bibitem{Casadio:2022ndh}
R.~Casadio, A.~Giusti, and J.~Ovalle, ``{Quantum Reissner-Nordstr\"om geometry:
  singularity and Cauchy horizon},'' {\em
  \href{https://arxiv.org/abs/2203.03252}{{\ttfamily arXiv:2203.03252
  [gr-qc]}}}.

\bibitem{Barcelo:2022gii}
C.~Barcel\'o, V.~Boyanov, R.~Carballo-Rubio, and L.~J. Garay, ``{Classical mass
  inflation vs semiclassical inner horizon inflation},'' {\em
  \href{https://arxiv.org/abs/2203.13539}{{\ttfamily arXiv:2203.13539
  [gr-qc]}}}.

\bibitem{Knorr:2022kqp}
B.~Knorr and A.~Platania, ``{Sifting quantum black holes through the principle
  of least action},'' {\em \href{https://arxiv.org/abs/2202.01216}{{\ttfamily
  arXiv:2202.01216 [hep-th]}}}.

\bibitem{Giddings:1992kn}
S.~B. Giddings and A.~Strominger, ``{Dynamics of extremal black holes},'' {\em
  \href{http://dx.doi.org/10.1103/PhysRevD.46.627} {Phys. Rev. D \textbf{46}
  (1992), 627-637}}.

\bibitem{Bardeen:1999px}
J.~M. Bardeen and G.~T. Horowitz, ``{The Extreme Kerr throat geometry: A Vacuum
  analog of AdS(2) x S**2},'' {\em
  \href{http://dx.doi.org/10.1103/PhysRevD.60.104030} {Phys. Rev. D \textbf{60}
  (1999), 104030}}.

\bibitem{Hartman:2008pb}
T.~Hartman, K.~Murata, T.~Nishioka, and A.~Strominger, ``{CFT Duals for Extreme
  Black Holes},'' {\em \href{http://dx.doi.org/10.1088/1126-6708/2009/04/019}
  {JHEP \textbf{04} (2009), 019}}.

\bibitem{Kunduri:2013gce}
H.~K. Kunduri and J.~Lucietti, ``{Classification of near-horizon geometries of
  extremal black holes},'' {\em \href{http://dx.doi.org/10.12942/lrr-2013-8}
  {Living Rev. Rel. \textbf{16} (2013), 8}}.

\bibitem{Maldacena:1998uz}
J.~M. Maldacena, J.~Michelson, and A.~Strominger, ``{Anti-de Sitter
  fragmentation},'' {\em \href{http://dx.doi.org/10.1088/1126-6708/1999/02/011}
  {JHEP \textbf{02} (1999), 011}}.

\bibitem{Almheiri:2014cka}
A.~Almheiri and J.~Polchinski, ``{Models of AdS$_{2}$ backreaction and
  holography},'' {\em \href{http://dx.doi.org/10.1007/JHEP11(2015)014} {JHEP
  \textbf{11} (2015), 014}}.

\bibitem{Almheiri:2016fws}
A.~Almheiri and B.~Kang, ``{Conformal Symmetry Breaking and Thermodynamics of
  Near-Extremal Black Holes},'' {\em
  \href{http://dx.doi.org/10.1007/JHEP10(2016)052} {JHEP \textbf{10} (2016),
  052}}.

\bibitem{Jackiw:1984je}
R.~Jackiw, ``{Lower Dimensional Gravity},'' {\em
  \href{http://dx.doi.org/10.1016/0550-3213(85)90448-1} {Nucl. Phys. B
  \textbf{252} (1985), 343-356}}.

\bibitem{Teitelboim:1983ux}
C.~Teitelboim, ``{Gravitation and Hamiltonian Structure in Two Space-Time
  Dimensions},'' {\em \href{http://dx.doi.org/10.1016/0370-2693(83)90012-6}
  {Phys. Lett. B \textbf{126} (1983), 41-45}}.

\bibitem{Grumiller:2002nm}
D.~Grumiller, W.~Kummer, and D.~V. Vassilevich, ``{Dilaton gravity in
  two-dimensions},'' {\em
  \href{http://dx.doi.org/10.1016/S0370-1573(02)00267-3} {Phys. Rept.
  \textbf{369} (2002), 327-430}}.

\bibitem{Hawking:1982dh}
S.~W. Hawking and D.~N. Page, ``{Thermodynamics of Black Holes in anti-De
  Sitter Space},'' {\em \href{http://dx.doi.org/10.1007/BF01208266} {Commun.
  Math. Phys. \textbf{87} (1983), 577}}.

\bibitem{Pavon:1991kh}
D.~Pavon, ``{Phase transition in Reissner-Nordstrom black holes},'' {\em
  \href{http://dx.doi.org/10.1103/PhysRevD.43.2495} {Phys. Rev. D \textbf{43}
  (1991), 2495-2497}}.

\bibitem{Witten:1998zw}
E.~Witten, ``{Anti-de Sitter space, thermal phase transition, and confinement
  in gauge theories},'' {\em
  \href{http://dx.doi.org/10.4310/ATMP.1998.v2.n3.a3} {Adv. Theor. Math. Phys.
  \textbf{2} (1998), 505-532}}.

\bibitem{Chamblin:1999hg}
A.~Chamblin, R.~Emparan, C.~V. Johnson, and R.~C. Myers, ``{Holography,
  thermodynamics and fluctuations of charged AdS black holes},'' {\em
  \href{http://dx.doi.org/10.1103/PhysRevD.60.104026} {Phys. Rev. D \textbf{60}
  (1999), 104026}}.

\bibitem{Wu:2000id}
X.~N. Wu, ``{Multicritical phenomena of Reissner-Nordstrom anti-de Sitter black
  holes},'' {\em \href{http://dx.doi.org/10.1103/PhysRevD.62.124023} {Phys.
  Rev. D \textbf{62} (2000), 124023}}.

\bibitem{Cadoni:2009xm}
M.~Cadoni, G.~D'Appollonio, and P.~Pani, ``{Phase transitions between
  Reissner-Nordstrom and dilatonic black holes in 4D AdS spacetime},'' {\em
  \href{http://dx.doi.org/10.1007/JHEP03(2010)100} {JHEP \textbf{03} (2010),
  100}}.

\bibitem{Kubiznak:2012wp}
D.~Kubiznak and R.~B. Mann, ``{P-V criticality of charged AdS black holes},''
  {\em \href{http://dx.doi.org/10.1007/JHEP07(2012)033} {JHEP \textbf{07}
  (2012), 033}}.

\bibitem{Rajagopal:2014ewa}
A.~Rajagopal, D.~Kubiz\v{n}\'ak, and R.~B. Mann, ``{Van der Waals black
  hole},'' {\em \href{http://dx.doi.org/10.1016/j.physletb.2014.08.054} {Phys.
  Lett. B \textbf{737} (2014), 277-279}}.

\bibitem{Mandal:2016anc}
A.~Mandal, S.~Samanta, and B.~R. Majhi, ``{Phase transition and critical
  phenomena of black holes: A general approach},'' {\em
  \href{http://dx.doi.org/10.1103/PhysRevD.94.064069} {Phys. Rev. D \textbf{94}
  (2016) no.6, 064069}}.

\bibitem{Li:2020nsy}
R.~Li, K.~Zhang, and J.~Wang, ``{Thermal dynamic phase transition of
  Reissner-Nordstr\"om Anti-de Sitter black holes on free energy landscape},''
  {\em \href{http://dx.doi.org/10.1007/JHEP10(2020)090} {JHEP \textbf{10}
  (2020), 090}}.

\bibitem{Chen:2014jwq}
P.~Chen, Y.~C. Ong, and D.-h. Yeom, ``{Black Hole Remnants and the Information
  Loss Paradox},'' {\em \href{http://dx.doi.org/10.1016/j.physrep.2015.10.007}
  {Phys. Rept. \textbf{603} (2015), 1-45}}.

\bibitem{Banerjee:2010qk}
R.~Banerjee, S.~K. Modak, and S.~Samanta, ``{Glassy Phase Transition and
  Stability in Black Holes},'' {\em
  \href{http://dx.doi.org/10.1140/epjc/s10052-010-1443-y} {Eur. Phys. J. C
  \textbf{70} (2010), 317-328}}.

\bibitem{Press:1971wr}
W.~H. Press, ``{Long Wave Trains of Gravitational Waves from a Vibrating Black
  Hole},'' {\em \href{http://dx.doi.org/10.1086/180849} {Astrophys. J. Lett.
  \textbf{170} (1971), L105-L108}}.

\bibitem{Ferrari:1984zz}
V.~Ferrari and B.~Mashhoon, ``{New approach to the quasinormal modes of a black
  hole},'' {\em \href{http://dx.doi.org/10.1103/PhysRevD.30.295} {Phys. Rev. D
  \textbf{30} (1984), 295-304}}.

\bibitem{Mashhoon:1985cya}
B.~Mashhoon, ``{Stability of charged rotating black holes in the eikonal
  approximation},'' {\em \href{http://dx.doi.org/10.1103/PhysRevD.31.290}
  {Phys. Rev. D \textbf{31} (1985) no.2, 290-293}}.

\bibitem{Cardoso:2008bp}
V.~Cardoso, A.~S. Miranda, E.~Berti, H.~Witek, and V.~T. Zanchin, ``{Geodesic
  stability, Lyapunov exponents and quasinormal modes},'' {\em
  \href{http://dx.doi.org/10.1103/PhysRevD.79.064016} {Phys. Rev. D \textbf{79}
  (2009), 064016}}.

\bibitem{Churilova:2019jqx}
M.~S. Churilova, ``{Analytical quasinormal modes of spherically symmetric black
  holes in the eikonal regime},'' {\em
  \href{http://dx.doi.org/10.1140/epjc/s10052-019-7146-0} {Eur. Phys. J. C
  \textbf{79} (2019) no.7, 629}}.

\bibitem{Schutz:1985km}
B.~F. Schutz and C.~M. Will, ``{BLACK HOLE NORMAL MODES: A SEMIANALYTIC
  APPROACH},'' {\em \href{http://dx.doi.org/10.1086/184453} {Astrophys. J.
  Lett. \textbf{291} (1985), L33-L36}}.

\bibitem{Yang:2012pj}
H.~Yang, F.~Zhang, A.~Zimmerman, D.~A. Nichols, E.~Berti, and Y.~Chen,
  ``{Branching of quasinormal modes for nearly extremal Kerr black holes},''
  {\em \href{http://dx.doi.org/10.1103/PhysRevD.87.041502} {Phys. Rev. D
  \textbf{87} (2013) no.4, 041502}}.

\bibitem{Frolov:2016pav}
V.~P. Frolov, ``{Notes on nonsingular models of black holes},'' {\em
  \href{http://dx.doi.org/10.1103/PhysRevD.94.104056} {Phys. Rev. D \textbf{94}
  (2016) no.10, 104056}}.

\bibitem{Molina:2021hgx}
M.~Molina and J.~R. Villanueva, ``{On the thermodynamics of the Hayward black
  hole},'' {\em \href{http://dx.doi.org/10.1088/1361-6382/abdd47} {Class.
  Quant. Grav. \textbf{38} (2021) no.10, 105002}}.

\bibitem{Ansoldi:2008jw}
S.~Ansoldi, ``{Spherical black holes with regular center: A Review of existing
  models including a recent realization with Gaussian sources},'' in {\em
  {Conference on Black Holes and Naked Singularities}}, 2 2008.

\bibitem{Casadio:2013hja}
R.~Casadio and A.~Orlandi, ``{Quantum Harmonic Black Holes},'' {\em
  \href{http://dx.doi.org/10.1007/JHEP08(2013)025} {JHEP \textbf{08} (2013),
  025}}.

\bibitem{Myung:2006mz}
Y.~S. Myung, Y.-W. Kim, and Y.-J. Park, ``{Thermodynamics and evaporation of
  the noncommutative black hole},'' {\em
  \href{http://dx.doi.org/10.1088/1126-6708/2007/02/012} {JHEP \textbf{02}
  (2007), 012}}.

\bibitem{Banerjee:2008du}
R.~Banerjee, B.~R. Majhi, and S.~K. Modak, ``{Noncommutative Schwarzschild
  Black Hole and Area Law},'' {\em
  \href{http://dx.doi.org/10.1088/0264-9381/26/8/085010} {Class. Quant. Grav.
  \textbf{26} (2009), 085010}}.

\bibitem{Nozari:2008rc}
K.~Nozari and S.~H. Mehdipour, ``{Hawking Radiation as Quantum Tunneling from
  Noncommutative Schwarzschild Black Hole},'' {\em
  \href{http://dx.doi.org/10.1088/0264-9381/25/17/175015} {Class. Quant. Grav.
  \textbf{25} (2008), 175015}}.

\end{thebibliography}
\bibliographystyle{ieeetr}
 
\end{document}